\documentclass[12pt]{article}
\usepackage{graphicx}
\usepackage{lscape}
\usepackage{epsfig}
\usepackage{citesort}
\usepackage{amssymb}
\usepackage{amsmath}
\usepackage{multirow}
\newcommand{\be}{\begin{equation}}
\newcommand{\ee}{\end{equation}}
\newcommand{\bea}{\begin{eqnarray}}
\newcommand{\eea}{\end{eqnarray}}
\newcommand{\nn}{\nonumber}

\def\R1{\varepsilon_1}
\def\E8{\varepsilon_8}

\def\ga{\gamma}

\def\lb{\Lambda_b}

\def\s1{\hat s}
\def\ds{\displaystyle}

\newcommand{\bd}{\begin{displaymath}}
\newcommand{\ed}{\end{displaymath}}

\newcommand{\f}{\frac}

\def\R1{\varepsilon_1}
\def\E8{\varepsilon_8}

\def\ga{\gamma}

\def\ds{\displaystyle}
\def\beq{\begin{equation}}
\def\eeq{\end{equation}}
\def\bea{\begin{eqnarray}}
\def\eea{\end{eqnarray}}
\def\beeq{\begin{eqnarray}}
\def\eeeq{\end{eqnarray}}

\def\vel{\left|}
\def\ver{\right|}
\def\nnb{\nonumber}
\def\ga{\left(}
\def\dr{\right)}

\def\rar{\rightarrow}
\def\nnb{\nonumber}

\def\ba{\begin{array}}
\def\ea{\end{array}}

\def\xis0{{\Xi^{*0}}}

\def\g5{\gamma_5}

\def\es{\!\!\! &=& \!\!\!}

\def\ar{&+& \!\!\!}
\def\ek{&-& \!\!\!}

\def\cp{&\times& \!\!\!}

\newcommand{\al}{\alpha_s}

\begin{document}

\title{ {\Large \textbf{Investigation of the  $\Lambda_b \rightarrow \Lambda \ell^+ \ell^-$
transition in universal extra dimension using  form factors from full QCD} } }
\author{\vspace{1cm} \\
%EndAName
{\small K. Azizi \thanks{%
e-mail: kazizi@dogus.edu.tr}\,\,, N. Kat{\i}rc{\i}\thanks{%
e-mail: nkatirci@dogus.edu.tr}} \\
{\small Physics Division, Faculty of Arts and Sciences, Do\u gu\c s
University} \\
{\small Ac{\i}badem-Kad{\i}k\"oy, 34722 Istanbul, Turkey} }
\date{}

\begin{titlepage}
\maketitle
\thispagestyle{empty}
\begin{abstract}
Using the related form factors from full QCD which recently are available, we provide a comprehensive analysis of the $\Lambda_b \rightarrow
\Lambda \ell^+ \ell^-$ transition in universal extra dimension model in the presence of a single universal extra dimension called the Applequist-Cheng-Dobrescu
model. In particular, we
analyze
  some related observables like  branching
ratio, forward-backward asymmetry, double lepton polarization asymmetries and
polarization of the $\Lambda$ baryon in terms of compactification
radius and corresponding form factors. We present the
sensitivity of these observables
  to the compactification parameter,
$1/R$ up
to $1/R=1000~GeV$. We also compare the results with those obtained using the form factors from heavy quark effective theory as well as
 the SM predictions.
\end{abstract}
~~~PACS number(s): 12.60-i,  13.30.-a, 13.30.Ce ,14.20.Mr
\end{titlepage}
%beyond the SM
%Kaluza-Klein theories
%Kaluza-Klein excitations (particle physics),
%Decays of baryons
%bottom baryons
%Sum rules
%hqet
\section{Introduction}
The Standard Model (SM) of particle physics describes all known
particles and their interactions except than gravity. The SM is the
only minimal model which is in perfect consistency with all
confirmed collider data despite it needs a missing ingredient, the
Higgs boson or something else to give masses to the elementary
particles. However, there are some problems such as  origin of the
matter in the universe, gauge and fermion mass hierarchy,  number of
generations, matter-antimatter asymmetry, unification, quantum
gravity and so on, which can not addressed by the SM. Such problems
show that 
   the SM can not be the ultimate theory of nature and it can
be considered as a low energy manifestation of some fundamental
theories.

Models with extra dimensions (ED)  \cite{Antoniadis1,Antoniadis2,arkani}  are among the most interesting
 candidates  beyond the SM to
overcome the aforementioned problems. A category of ED which allows
the SM fields (both gauge bosons and fermions) propagate  in the
extra dimensions called the universal extra dimension (UED) model.
The most simple example of the UED model also, where just a single
universal extra dimension compactified on a circle of radius $R$ is
considered, is called the Appelquist, Cheng and Dobrescu (ACD) model
\cite{acdd}. The  radius $R$ is the extra
parameter that causes the difference between SM and
its beyond. The particles with momentum in extra dimension are
called Kaluza-Klein (KK) particles. The mass of KK particles and their interaction with  themselves as well as with the SM particles are described in terms of  compactification scale, $1/R$.
One of the important property of the model is the conservation
of KK parity that guarantees the absence of tree level KK contributions to low energy processes occurring  at scales very smaller than the compactification scale \cite{Buras:2002ej} 
(for more information about the 
model see also \cite{R7623,R7624}). The 
flavor changing neutral current (FCNC) transition of $\Lambda_b \rightarrow \Lambda \ell^+ \ell^-$, which may be in the future program of the LHCb to study, lies in this scale. This transition is proceed via the FCNC transition 
of $b \rightarrow s \ell^+ \ell^-$ at loop level in SM via  electroweak penguin and weak box diagrams, which  are sensitive to new physics contributions. Looking for SUSY
particles \cite{10susy}, light dark matter \cite{11darkmatter},  probable fourth generation of the quarks, and also KK modes (extra dimensions
) \cite{Buras:2002ej}
 is possible by investigating such
loop level transitions.

The aim of the paper is to
find the effects of the KK modes on various observables related to the $\Lambda_{b}\rar \Lambda \ell^+\ell^- $ transition. These observables are total decay rate, branching
ratio, forward-backward asymmetry, double lepton polarization asymmetries and
polarization of the $\Lambda$ baryon. 
 We  analyze
 these observables in terms of the corresponding form factors as well as the compactification factor.  From the electroweak precision tests, the lower limit for $1/R$ is obtained as $%
250~GeV$  if $M_h\geq250~GeV$ expressing larger KK contributions to the low energy FCNC processes like, $\Lambda_{b}\rar \Lambda \ell^+\ell^- $,
 and $300~GeV$  if $M_h\leq250~GeV$, respectively \cite{acdd,T. Appelquist}. 
%Analyses of the $B\rightarrow X_s\gamma$ transition  and also anomalous magnetic moment
  %have shown that the experimental data are in a good agreement with the ACD model if $1/R\geq 300~GeV$ \cite{ikiuc}.
 We will consider the   $1/R$ from $200~GeV$ up
to $1000~GeV$. We will use also the form factors  obtained both using QCD sum rules
 in full theory, which they recently are available \cite{kazizi} and also those obtained  in heavy quark effective theory (HQET) \cite{huang}. Using the values of the form factors, we present the
sensitivity of these observables
  to the compactification parameter,
$1/R$. Note that, using the form factors obtained in HQET, the transitions, $\Lambda_b\rightarrow\Lambda \nu\bar\nu$ and $\Lambda_b\rightarrow\Lambda \gamma$  \cite{colangelo}, 
 $\Lambda_b\rightarrow\Lambda l^+l^-$ \cite{R7601},  
$\Lambda_b\rightarrow\Lambda \gamma$ and  $\Lambda_b\rightarrow\Lambda l^+l^-$
\cite{wang} have been analyzed in the same framework. The ACD model also has been applied to investigate some $B$ and $K$ mesons decays in \cite{Buras:2002ej,R7624,R7623,colangelobey,Ishtiaq,Asif,Zeynali}.

The layout of the paper is as follows. In  next section, we  introduce   responsible Hamiltonian and present   Wilson coefficients  in  UED  model. We also  present the 
 transition matrix elements in terms of  form factors responsible for $\Lambda_{b}\rar \Lambda \ell^+\ell^- $ transition in this section. In section 3, 
we analyze
the branching ratio,
forward-backward asymmetry, double lepton polarization asymmetries and polarization of
the $\Lambda$ baryon in terms of the compactification factor, $1/R$. In this section, using the form factors both from full theory and HQET, we also  compare our results obtained both in the  UED and SM models for each observable and
discuss the results. Finally, we will present our concluding remark in section 4.

\section{Effective Hamiltonian, Transition Matrix elements and Form Factors Responsible for $\Lambda_b \rightarrow
\Lambda \ell^+ \ell^-$ }
\subsection{Effective Hamiltonian}
The $\Lambda_b \rightarrow
\Lambda \ell^+ \ell^-$  transition is governed by the  FCNC transition of the $b \rar s l^+ l^- $ at quark level and is described by the following   effective Hamiltonian:

 \bea \label{e8401} {\cal H}^{eff} &=& {G_F \alpha_{em} V_{tb}
V_{ts}^\ast \over 2\sqrt{2} \pi} \Bigg[ C_9^{eff} 
\bar{s}\gamma_\mu (1-\gamma_5) b \, \bar{\ell} \gamma^\mu \ell +
C_{10}  \bar{s} \gamma_\mu (1-\gamma_5) b \, \bar{\ell}
\gamma^\mu
\gamma_5 \ell \nnb \\
&-&  2 m_b C_7^{eff}  {1\over q^2} \bar{s} i \sigma_{\mu\nu}
(1+\gamma_5) b \, \bar{\ell} \gamma^\mu \ell \Bigg]~, \eea

where $\alpha_{em}$  is the fine structure
constant at Z mass scale, $G_F$ is the Fermi constant,  $V_{ij}$ are elements of the
Cabibbo-Kobayashi-Maskawa (CKM) matrix and 
$C_7^{eff}$, $C_9^{eff}$ and $C_{10}$ are the Wilson coefficients. These coefficients  are the main source of the deviation of  the ACD model results for the observables from the  SM models predictions.
These coefficients are expressed in terms of  some periodic
functions, $F(x_{t},1/R)$ with $x_{t}=\frac{m_{t}^{2}}{M_{W}^{2}}$ and $m_t$ being 
the top quark mass. The mass of the KK particles are represented in terms of the zero modes 
corresponding to the ordinary SM  particles and an extra part coming from the ACD model, i.e., $m_n^2=m_0^2+\frac{n^2}{R^2}$. Similar to the mass of the KK particles,  the functions, $F(x_{t},1/R)$ 
are also  described in terms of the corresponding SM functions, $F_0 (x_t )$ and   functions  in terms of the
compactification factor, $1/R$,
\bea F(x_t,1/R)=F_0(x_t)+\sum_{n=1}^{\infty}F_n(x_t,x_n),
\label{functions} \eea where $x_n=\displaystyle{m_n^2 \over
m_W^2}$ and $m_n=\displaystyle{n \over R}$. The Glashow-Illiopoulos-Maiani (GIM) mechanism undertakes the finiteness of the functions, $F(x_t,1/R)$ and fulfills the 
condition, $F(x_t,1/R)\rar F_0(x_t)$, when $R\rar 0$. As far as the compactification factor, $1/R$ is recorded in  order of a few hundreds of $GeV$, these functions and as a result, the Wilson coefficients and
considered  observables
 differ considerably from the SM predictions. In the following, we present the explicit expressions of the  Wilson coefficients as well as their numerical values from $1/R=200~GeV$ up to $1/R=1000~GeV$
 in ACD model  ( see also \cite{Buras:2002ej,R7623,R7624}).

In ACD model with a single universal extra dimension, the  $C_7^{eff}(1/R)$ in leading log approximation
is written as (see also \cite{R7626}):
  \bea
\label{e7603} C_7^{eff}(\mu_b, 1/R) \es \eta^{\frac{16}{23}}
C_7(\mu_W, 1/R)+ \frac{8}{3} \left( \eta^{\frac{14}{23}}
-\eta^{\frac{16}{23}} \right) C_8(\mu_W, 1/R)+C_2 (\mu_W)
\sum_{i=1}^8 h_i \eta^{a_i}~, \nnb\\ \eea 
where the first and second arguments show the scale and  dependency on the compactification parameters, $1/R$, respectively and,
\bea
 C_2(\mu_W)=1~,~~ C_7(\mu_W, 1/R)=-\frac{1}{2}
D^\prime(x_t,1/R)~,~~ C_8(\mu_W, 1/R)=-\frac{1}{2}
E^\prime(x_t,1/R)~ . \eea
The functions,  $D^\prime (x_t,1/R)$  and $E^\prime (x_t,1/R)$ are given as:
\bea
D^\prime (x_t,1/R)=D^\prime_0(x_t)+\sum_{n=1}^{\infty}D^\prime_n(x_t,x_n),~~~~~E^\prime (x_t,1/R)=E^\prime_0(x_t)+\sum_{n=1}^{\infty}E^\prime_n(x_t,x_n)~, \eea
where,
 \bea \label{e7604} D^\prime_0(x_t) \es - \frac{(8
x_t^3+5 x_t^2-7 x_t)}{12 (1-x_t)^3}
+ \frac{x_t^2(2-3 x_t)}{2(1-x_t)^4}\ln x_t~, \\ \nnb \\
\label{e7605} E^\prime_0(x_t) \es - \frac{x_t(x_t^2-5 x_t-2)}{4
(1-x_t)^3} +
\frac{3 x_t^2}{2 (1-x_t)^4}\ln x_t~, 
\eea
 and
 \bea \label{e7608}
\sum_{n=1}^{\infty}D^\prime_n(x_t,x_n) \es
\frac{x_t[37 - x_t(44+17 x_t)]}{72 (x_t-1)^3} \nnb \\
\ar \frac{\pi m_W R}{12} \Bigg[ \int_0^1 dy \, (2 y^{1/2}+7
y^{3/2}+3 y^{5/2}) \, \coth (\pi m_WR \sqrt{y}) \nnb \\
\ek \frac{x_t (2-3 x_t) (1+3 x_t)}{(x_t-1)^4}J(R,-1/2)\nnb \\
\ek \frac{1}{(x_t-1)^4} \{ x_t(1+3 x_t)+(2-3 x_t)
[1-(10-x_t)x_t] \} J(R, 1/2)\nnb \\
\ek \frac{1}{(x_t-1)^4} [ (2-3 x_t)(3+x_t) + 1 - (10-x_t) x_t]
J(R, 3/2)\nnb \\
\ek \frac{(3+x_t)}{(x_t-1)^4} J(R,5/2)\Bigg]~, \\ \nnb \\
\label{e7609} \sum_{n=1}^{\infty}E^\prime_n(x_t,x_n)\es
\frac{x_t[17+(8-x_t)x_t]}
{24 (x_t-1)^3} \nnb \\
\ar \frac{\pi m_W R}{4} \Bigg[\int_0^1 dy \, (y^{1/2}+
2 y^{3/2}-3 y^{5/2}) \, \coth (\pi m_WR \sqrt{y}) \nnb \\
\ek {x_t(1+3 x_t) \over (x_t-1)^4}J(R,-1/2)\nnb \\
\ar \frac{1}{(x_t-1)^4} [ x_t(1+3 x_t) - 1 + (10-x_t)x_t] J(R, 1/2)\nnb \\
\ek \frac{1}{(x_t-1)^4} [(3+x_t)-1+(10-x_t)x_t) ]J(R, 3/2)\nnb \\
\ar{(3+x_t) \over  (x_t-1)^4} J(R,5/2)\Bigg]~, \eea with, \bea
\label{e76010} J(R,\alpha)=\int_0^1 dy \, y^\alpha \left[ \coth
(\pi m_W R \sqrt{y})-x_t^{1+\alpha} \coth(\pi m_t R \sqrt{y})
\right]~. \eea
The remaining parameters in Eq. (\ref{e7603}) are defined as:
 \bea \eta \es
\frac{\alpha_s(\mu_W)} {\alpha_s(\mu_b)}~,\eea
 \bea
\alpha_s(x)=\frac{\alpha_s(m_Z)}{1-\beta_0\frac{\alpha_s(m_Z)}{2\pi}\ln(\frac{m_Z}{x})},\eea
where in fifth dimension, $\alpha_s(m_Z)=0.118$ and $\beta_0=\frac{23}{3}$. The coefficients $a_i$ and $h_i$ are given as  \cite{R7627,R762777}:
   \be\frac{}{}
   \label{klar}
\begin{array}{rrrrrrrrrl}
a_i = (\!\! & \f{14}{23}, & \f{16}{23}, & \f{6}{23}, & -
\f{12}{23}, &
0.4086, & -0.4230, & -0.8994, & 0.1456 & \!\!)  \vspace{0.1cm}, \\
h_i = (\!\! & 2.2996, & - 1.0880, & - \f{3}{7}, & - \f{1}{14}, &
-0.6494, & -0.0380, & -0.0186, & -0.0057 & \!\!).
\end{array}
\ee

 The  Wilson coefficient $C_{10}$ is given  as:
 \bea
\label{e7616} C_{10}(1/R)= - \frac{Y(x_t,1/R)}{\sin^2 \theta_W}~, \eea
where, $\sin^2\theta_W= 0.23$ and,
 \bea
\label{e7612}
Y(x_t,1/R) \es Y_0(x_t)+\sum_{n=1}^\infty C_n(x_t,x_n)~, 
\eea with, \bea
\label{e7613} Y_0(x_t) \es \frac{x_t}{8} \left[ \frac{x_t -4}{x_t
-1}+\frac{3 x_t}{(x_t-1)^2} \ln x_t \right]~, \eea  and,  \bea
\label{e7615} \sum_{n=1}^\infty C_n(x_t,x_n) = \frac{x_t(7-x_t)}{
16 (x_t-1)} - \frac{\pi m_W R x_t}{16 (x_t-1)^2}
\left[3(1+x_t)J(R,-1/2)+(x_t-7)J(R,1/2) \right]~.\nnb\\ \eea 
Finally, we consider the Wilson coefficient,
$C_9^{eff}$.  It can be written  in leading log
approximation
 as \cite{R7627,R762777}:
 \bea \label{C9eff}
C_9^{eff}(\hat{s}',1/R) & = & C_9^{NDR}(1/R)\eta(\hat s') + h(z, \hat s')\left( 3
C_1 + C_2 + 3 C_3 + C_4 + 3
C_5 + C_6 \right) \nonumber \\
& & - \f{1}{2} h(1, \hat s') \left( 4 C_3 + 4 C_4 + 3
C_5 + C_6 \right) \nonumber \\
& & - \f{1}{2} h(0, \hat s') \left( C_3 + 3 C_4 \right)
+ \f{2}{9} \left( 3 C_3 + C_4 + 3 C_5 +
C_6 \right), \eea

where, $\hat{s}'=\frac{q^2}{m_b^2}$ with $4m_l^2\leq q^2\leq(m_{\Lambda_b}-m_\Lambda)^2$ and, 
\bea \label{C9tilde}C_9^{NDR}(1/R) & = & P_0^{NDR} +
\f{Y(x_t)}{\sin^2\theta_W} -4 Z(x_t) + P_E E(x_t), \eea 
 here, NDR stands for the naive dimensional regularization scheme. We neglect the  last term in this equation since 
 due to the order of $P_E$, the contribution of this term is negligibly small. The  $P_0^{NDR}=2.60 \pm 0.25$ \cite{R7627,R762777} and the function,
 $Z(x_t,1/R)$  is defined as:
\bea \label{e7612}
 Z(x_t,1/R) \es Z_0(x_t)+\sum_{n=1}^\infty
C_n(x_t,x_n)~, \eea where, \bea Z_0(x_t) \es \frac{18 x_t^4-163
x_t^3+259 x_t^2 -108 x_t}{144 (x_t-1)^3} \nnb +\left[\frac{32
x_t^4-38 x_t^3-15 x_t^2+18 x_t}{72
(x_t-1)^4} - \frac{1}{9}\right] \ln x_t .\\ \nnb \\
\eea 
In Eq. (\ref{C9eff}),
\bea \eta(\hat s') & = & 1 +
\f{\al(\mu_b)}{\pi}\, \omega(\hat s'), \eea
with,
 \bea \label{omega}
\omega(\hat s') & = & - \f{2}{9} \pi^2 - \f{4}{3}\mbox{Li}_2(\hat s') -
\f{2}{3}
\ln \hat s' \ln(1-\hat s') - \f{5+4\hat s'}{3(1+2\hat s')}\ln(1-\hat s') - \nonumber \\
& &  \f{2 \hat s' (1+\hat s') (1-2\hat s')}{3(1-\hat s')^2 (1+2\hat s')} \ln \hat s' + \f{5+9\hat s'-6\hat s'^2}{6
(1-\hat s') (1+2\hat s')}, \eea 
 and at $\mu_b$ scale, \bea \label{coeffs} C_j=\sum_{i=1}^8 k_{ji}
\eta^{a_i} \qquad (j=1,...6) \vspace{0.2cm} \eea where $k_{ji}$
 are given as:
\be\frac{}{}
   \label{klar}
\begin{array}{rrrrrrrrrl}
k_{1i} = (\!\! & 0, & 0, & \f{1}{2}, & - \f{1}{2}, &
0, & 0, & 0, & 0 & \!\!),  \vspace{0.1cm} \\
k_{2i} = (\!\! & 0, & 0, & \f{1}{2}, &  \f{1}{2}, &
0, & 0, & 0, & 0 & \!\!),  \vspace{0.1cm} \\
k_{3i} = (\!\! & 0, & 0, & - \f{1}{14}, &  \f{1}{6}, &
0.0510, & - 0.1403, & - 0.0113, & 0.0054 & \!\!),  \vspace{0.1cm} \\
k_{4i} = (\!\! & 0, & 0, & - \f{1}{14}, &  - \f{1}{6}, &
0.0984, & 0.1214, & 0.0156, & 0.0026 & \!\!),  \vspace{0.1cm} \\
k_{5i} = (\!\! & 0, & 0, & 0, &  0, &
- 0.0397, & 0.0117, & - 0.0025, & 0.0304 & \!\!) , \vspace{0.1cm} \\
k_{6i} = (\!\! & 0, & 0, & 0, &  0, &
0.0335, & 0.0239, & - 0.0462, & -0.0112 & \!\!).  \vspace{0.1cm} \\
\end{array}
\ee
The remaining functions in Eq. (\ref{C9eff}) are also given as:
 \bea \label{phasespace} h(y,
\hat s') & = & -\f{8}{9}\ln\f{m_b}{\mu_b} - \f{8}{9}\ln y +
\f{8}{27} + \f{4}{9} x \\
& & - \f{2}{9} (2+x) |1-x|^{1/2} \left\{
\begin{array}{ll}
\left( \ln\left| \f{\sqrt{1-x} + 1}{\sqrt{1-x} - 1}\right| - i\pi
\right), &
\mbox{for } x \equiv \f{4z^2}{\hat s'} < 1 \nonumber \\
2 \arctan \f{1}{\sqrt{x-1}}, & \mbox{for } x \equiv \f {4z^2}{\hat
s'} > 1,
\end{array}
\right. \\
\eea 
where   $y=1$ or $y=z=\frac{m_c}{m_b}$ and,
\bea h(0, \hat s') & = & \f{8}{27}
-\f{8}{9} \ln\f{m_b}{\mu_b} - \f{4}{9} \ln \hat s' + \f{4}{9}
i\pi.\eea
 
 Numerical results   show that the Wilson coefficients in UED 
differ from their SM values, considerably. In particular, the  $C_{10}$ is enhanced and the $C_7^{eff}$ is suppressed (for more details see \cite{R7626,R7627,R762777}).

\subsection{Transition Matrix Elements and Form Factors}
The decay 
amplitude of the $\Lambda_b \rightarrow
\Lambda \ell^+ \ell^-$  is obtained sandwiching  the aforementioned effective  Hamiltonian between
the initial and final baryonic states. As a result, the transition
matrix elements,  $\langle
\Lambda(p) \mid  \bar s \gamma_\mu (1-\gamma_5) b \mid \Lambda_b(p+q)\rangle $ and  $\langle \Lambda(p)\mid \bar s i \sigma_{\mu\nu}q^{\nu} (1+ \gamma_5)
b \mid \Lambda_b(p+q)\rangle $ are appeared. In full theory of QCD, they can be
parametrized  in terms of  twelve   form factors, $f_{i}$,
$g_{i}$, $f^T_{i}$ and $g^T_{i}$ ($i$ running from $1$ to $3$) in
the following manner:
\bea\label{matrixel1a}
\langle
\Lambda(p) \mid  \bar s \gamma_\mu (1-\gamma_5) b \mid \Lambda_b(p+q)\rangle\es
\bar {u}_\Lambda(p) \Big[\gamma_{\mu}f_{1}(q^{2})+{i}
\sigma_{\mu\nu}q^{\nu}f_{2}(q^{2}) + q^{\mu}f_{3}(q^{2}) \nnb \\
\ek \gamma_{\mu}\gamma_5
g_{1}(q^{2})-{i}\sigma_{\mu\nu}\gamma_5q^{\nu}g_{2}(q^{2})
- q^{\mu}\gamma_5 g_{3}(q^{2})
\vphantom{\int_0^{x_2}}\Big] u_{\Lambda_{b}}(p+q)~,\nnb \\
\eea
\bea\label{matrixel1b}
\langle \Lambda(p)\mid \bar s i \sigma_{\mu\nu}q^{\nu} (1+ \gamma_5)
b \mid \Lambda_b(p+q)\rangle \es\bar{u}_\Lambda(p)
\Big[\gamma_{\mu}f_{1}^{T}(q^{2})+{i}\sigma_{\mu\nu}q^{\nu}f_{2}^{T}(q^{2})+
q^{\mu}f_{3}^{T}(q^{2}) \nnb \\
\ar \gamma_{\mu}\gamma_5
g_{1}^{T}(q^{2})+{i}\sigma_{\mu\nu}\gamma_5q^{\nu}g_{2}^{T}(q^{2})
+ q^{\mu}\gamma_5 g_{3}^{T}(q^{2})
\vphantom{\int_0^{x_2}}\Big] u_{\Lambda_{b}}(p+q)~.\nnb \\
\eea
 These form factors have  been recently calculated in \cite{kazizi} using light cone QCD sum rules in
full theory. 

On the other hand, the aforesaid transition matrix elements  in HQET
 is defined in terms of only two form factors, $F_1$ and $F_2$ as
\cite{27alievozpineci,28Mannel}:
 \bea\label{matrixel1111} \langle
\Lambda(p) \mid \bar s\Gamma b\mid \Lambda_b(p+q)\rangle \es
\bar{u}_\Lambda(p)[F_1(q^2)+\not\!vF_2(q^2)]\Gamma
u_{\Lambda_b}(p+q), \eea where $\Gamma$ denotes   any 
Dirac matrices,
 $\not\!v=({\not\!p}+{\not\!q})/m_{\Lambda_{b}}$ and   the form factors, $F_1(q^2)$ and $F_2(q^2)$ have been calculated in \cite{huang}.
Comparing the definitions of the transition matrix elements both in full and HQET theories, one can easily find relations among the form factors mentioned above (see  \cite{kazizi,chen,30ozpineci}).
\section{ Some Observables Related to the $\Lambda_{b}\rightarrow \Lambda \ell^{+}\ell^{-}$}
\subsection{Branching Ratio}
Using the decay amplitude discussed in the previous section, the angular and $1/R$ 
dependent differential decay rate can be written as (see  \cite{R7601,savcibey,Giri}):

\bea \frac{d\Gamma}{d\hat sdz}(z,\hat s,1/R) = \frac{G_F^2\alpha^2_{em}
m_{\Lambda_b}}{16384 \pi^5}| V_{tb}V_{ts}^*|^2 v \sqrt{\lambda} \,
\Bigg[{\cal T}_0(\hat s,1/R)+{\cal T}_1(\hat s,1/R) z +{\cal
T}_2(\hat s,1/R) z^2\Bigg]~, \nnb\\ \label{rate} \eea
where
$z=\cos\theta$,
 $\theta$ being the angle between the momenta
of $\Lambda_b$ and $\ell^-$ in the
center of mass of leptons,
$\lambda=\lambda(1,r,\hat s)=1+r^2+\hat s^2-2r-2\hat s-2r\hat s$, $r= m^2_{\Lambda}/m^2_{\Lambda_b}$ and
$v=\sqrt{1-\frac{4 m_\ell^2}{q^2}}$.
 The functions, ${\cal T}_0(\hat s,1/R)$,  ${\cal T}_1(\hat s,1/R)$ and
${\cal T}_2(\hat s,1/R)$ are given as (see also \cite{kazizi}):
 \bea {\cal T}_0(\hat s,1/R) \es 32 m_\ell^2
m_{\Lambda_b}^4 \hat s (1+r-\hat s) \ga \vel D_3 \ver^2 +
\vel E_3 \ver^2 \dr \nnb \\
\ar 64 m_\ell^2 m_{\Lambda_b}^3 (1-r-\hat s) \, \mbox{\rm Re} [D_1^\ast
E_3 + D_3
E_1^\ast] \nnb \\
\ar 64 m_{\Lambda_b}^2 \sqrt{r} (6 m_\ell^2 - m_{\Lambda_b}^2 \hat s)
{\rm Re} [D_1^\ast E_1] \nnb \\
\ar 64 m_\ell^2 m_{\Lambda_b}^3 \sqrt{r} \Big( 2 m_{\Lambda_b} \hat s
{\rm Re} [D_3^\ast E_3] + (1 - r + \hat s)
{\rm Re} [D_1^\ast D_3 + E_1^\ast E_3]\Big) \nnb \\
\ar 32 m_{\Lambda_b}^2 (2 m_\ell^2 + m_{\Lambda_b}^2 \hat s) \Big\{ (1
- r + \hat s) m_{\Lambda_b} \sqrt{r} \,
\mbox{\rm Re} [A_1^\ast A_2 + B_1^\ast B_2] \nnb \\
\ek m_{\Lambda_b} (1 - r - \hat s) \, \mbox{\rm Re} [A_1^\ast B_2 +
A_2^\ast B_1] - 2 \sqrt{r} \Big( \mbox{\rm Re} [A_1^\ast B_1] +
m_{\Lambda_b}^2 \hat s \,
\mbox{\rm Re} [A_2^\ast B_2] \Big) \Big\} \nnb \\
\ar 8 m_{\Lambda_b}^2 \Big\{ 4 m_\ell^2 (1 + r - \hat s) +
m_{\Lambda_b}^2 \Big[(1-r)^2 - \hat s^2 \Big]
\Big\} \ga \vel A_1 \ver^2 +  \vel B_1 \ver^2 \dr \nnb \\
\ar 8 m_{\Lambda_b}^4 \Big\{ 4 m_\ell^2 \Big[ \lambda + (1 + r -
\hat s) \hat s \Big] + m_{\Lambda_b}^2 \hat s \Big[(1-r)^2 - \hat s^2 \Big]
\Big\} \ga \vel A_2 \ver^2 +  \vel B_2 \ver^2 \dr \nnb \\
\ek 8 m_{\Lambda_b}^2 \Big\{ 4 m_\ell^2 (1 + r - \hat s) -
m_{\Lambda_b}^2 \Big[(1-r)^2 - \hat s^2 \Big]
\Big\} \ga \vel D_1 \ver^2 +  \vel E_1 \ver^2 \dr \nnb \\
\ar 8 m_{\Lambda_b}^5 \hat s v^2 \Big\{ - 8 m_{\Lambda_b} \hat s \sqrt{r}\,
\mbox{\rm Re} [D_2^\ast E_2] +
4 (1 - r + \hat s) \sqrt{r} \, \mbox{\rm Re}[D_1^\ast D_2+E_1^\ast E_2]\nnb \\
\ek 4 (1 - r - \hat s) \, \mbox{\rm Re}[D_1^\ast E_2+D_2^\ast E_1] +
m_{\Lambda_b} \Big[(1-r)^2 -\hat s^2\Big] \ga \vel D_2 \ver^2 + \vel
E_2 \ver^2\dr \Big\},\nnb \\
\eea
\bea {\cal T}_1(\hat s,1/R) &=& -16
m_{\lb}^4\s1 v \sqrt{\lambda}
\Big\{ 2 Re(A_1^* D_1)-2Re(B_1^* E_1)\nn\\
&+& 2m_{\lb}
Re(B_1^* D_2-B_2^* D_1+A_2^* E_1-A_1^*E_2)\Big\}\nn\\
&+&32 m_{\lb}^5 \s1~ v \sqrt{\lambda} \Big\{
m_{\lb} (1-r)Re(A_2^* D_2 -B_2^* E_2)\nn\\
&+& \sqrt{r} Re(A_2^* D_1+A_1^* D_2-B_2^*E_1-B_1^* E_2)\Big\}\;,
\eea
 \bea {\cal T}_2(\hat s,1/R)\es - 8 m_{\Lambda_b}^4 v^2 \lambda \ga
\vel A_1 \ver^2 + \vel B_1 \ver^2 + \vel D_1 \ver^2
+ \vel E_1 \ver^2 \dr \nnb \\
\ar 8 m_{\Lambda_b}^6 \hat s v^2 \lambda \Big( \vel A_2 \ver^2 + \vel
B_2 \ver^2 + \vel D_2 \ver^2 + \vel E_2 \ver^2  \Big)~, \eea where,
 \bea \label{a9} A_1 \es \frac{1}{q^2}\ga
f_1^T+g_1^T \dr \ga -2 m_b C_7^{eff}(1/R)\dr + \ga f_1-g_1 \dr C_9^{eff}(\hat s,1/R) \nnb \\
A_2 \es A_1 \ga 1 \rar 2 \dr ~,\nnb \\
A_3 \es A_1 \ga 1 \rar 3 \dr ~,\nnb \\
B_1 \es A_1 \ga g_1 \rar - g_1;~g_1^T \rar - g_1^T \dr ~,\nnb \\
B_2 \es B_1 \ga 1 \rar 2 \dr ~,\nnb \\
B_3 \es B_1 \ga 1 \rar 3 \dr ~,\nnb \\
D_1 \es \ga f_1-g_1 \dr C_{10}(1/R) ~,\nnb \\
D_2 \es D_1 \ga 1 \rar 2 \dr ~, \\
D_3 \es D_1 \ga 1 \rar 3 \dr ~,\nnb \\
E_1 \es D_1 \ga g_1 \rar - g_1 \dr ~,\nnb \\
E_2 \es E_1 \ga 1 \rar 2 \dr ~,\nnb \\
E_3 \es E_1 \ga 1 \rar 3 \dr ~,
 \eea 
and the relation between the $\hat s'$ used in the previous section and $\hat s$ in the present section is: $\hat s'=\frac{\hat s~ m_{\Lambda_b}^2}{m_b^2}$.
Integrating out the angular
dependent differential decay rate,  the following dilepton mass spectrum is obtained:
 \bea
\frac{d\Gamma}{d \hat s} (\hat s,1/R)= \frac{G_F^2\alpha^2_{em} m_{\Lambda_b}}{8192
\pi^5}| V_{tb}V_{ts}^*|^2 v \sqrt{\lambda} \, \Bigg[{{\cal T}_0(\hat s,1/R)
+\frac{1}{3} {\cal T}_2(\hat s,1/R)}\Bigg]~. \label{rate} \eea
Integrating also the above equation over $\hat s$ in the allowed physical region, $\frac{4 m_\ell^2}{m_{\Lambda_b}^2}\le\hat s \le (1-\sqrt{r})^2$, one can obtain the   $1/R$ dependent total decay width. Multiplying  the 
total decay rate to the lifetime of the $\Lambda_b$ baryon, we obtain the $1/R$ dependent branching ratio. Using the numerical values, $m_t=167~GeV$, $m_W=80.4~GeV$, $m_Z=91~GeV$, $m_b=4.8~GeV$,
 $m_c=1.46~GeV$, $\mu_b=5~GeV$,
 $\mu_W=80.4~GeV$,
$| V_{tb}V_{ts}^\ast|=0.041$, $G_F = 1.17 \times 10^{-5}~ GeV^{-2}$, $\alpha_{em}=\frac{1}{137}$,
  $\tau_{\Lambda_b}=1.383\times 10^{-12}~s$, $m_\Lambda =
1.115~GeV$, $m_{\Lambda_{b}} = 5.620~ GeV$ \cite{33Yao:2006px}, $m_e = 0.51~MeV$,  $m_\mu = 0.1056~GeV$ and   $m_\tau = 1.771~GeV$,  we present the dependence of  branching ratios on compactification factor, $1/R$
 in Fig. 1.
\begin{figure}
\label{fig1}
\centering
\begin{tabular}{ccc}
\epsfig{file=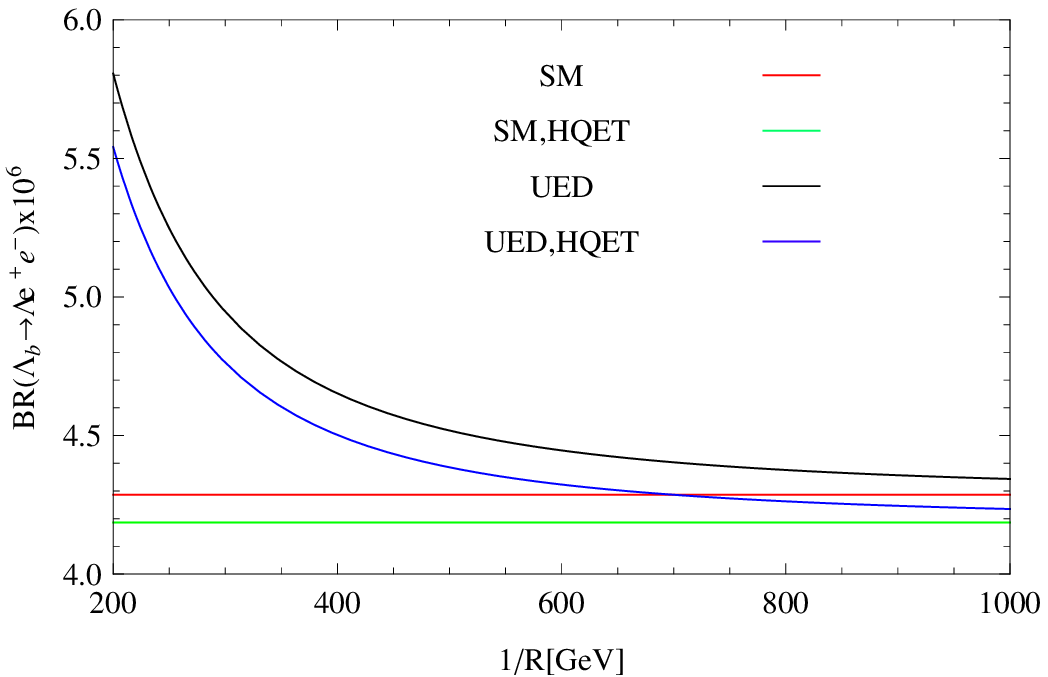,width=0.33\linewidth,clip=} &
\epsfig{file=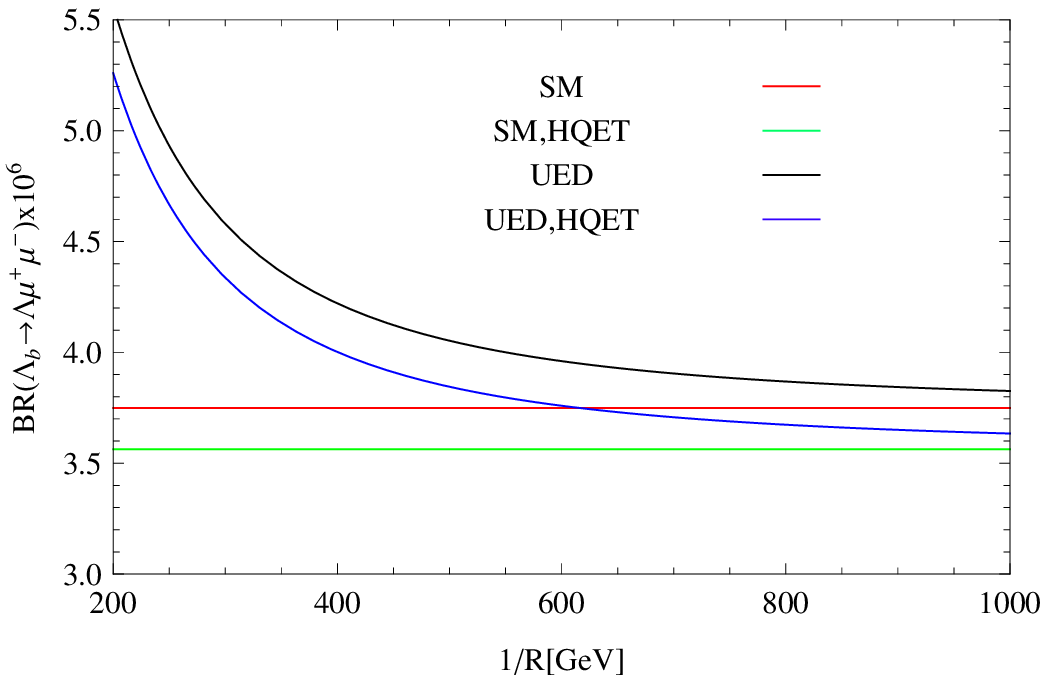,width=0.33\linewidth,clip=} &
\epsfig{file=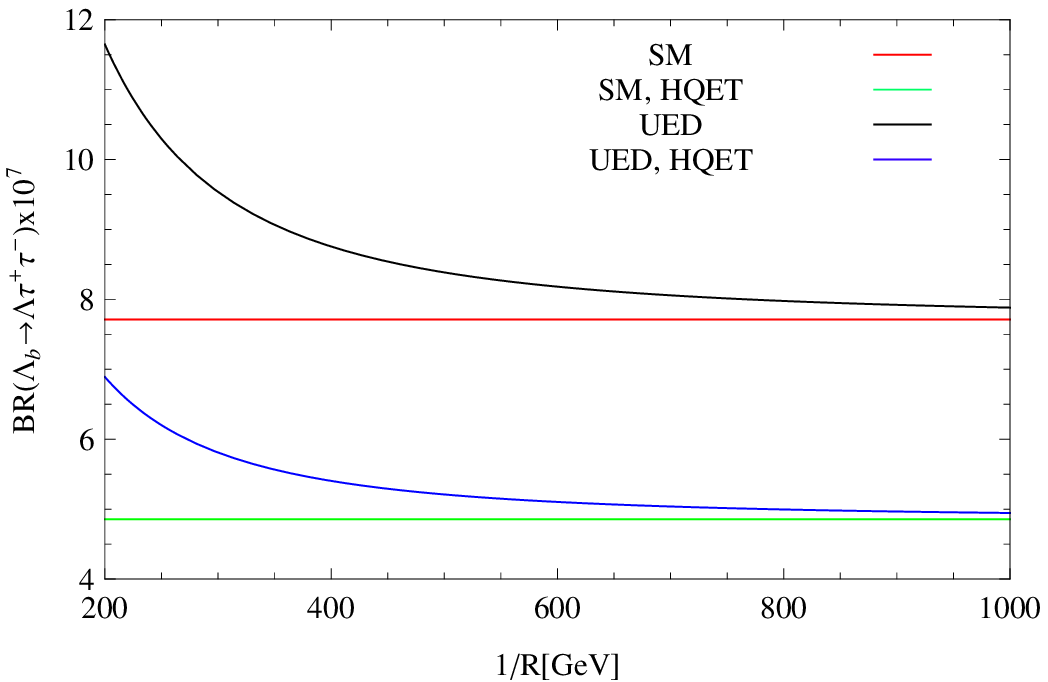,width=0.33\linewidth,clip=}
\end{tabular}
\caption{The dependence of  branching ratios on compactification factor, $1/R$.}
\end{figure}
 From this figure, we deduce the following results:
\begin{itemize}
 \item There is  considerable discrepancy between the predictions of the ACD and SM models for  low  values of the compactification factor, $1/R$. 
As  $1/R$ increases, this difference tends to  diminish 
so that for higher values of   $1/R$ ($1/R\simeq1000~GeV$), the predictions of ACD  become very close to the  results of  SM . Such  discrepancy at low values of $1/R$ can be 
 considered as a signal for the  existence of extra dimensions.
\item As it is expected, an increase in the lepton mass  ends up in a decrease in the branching ratio. 
\item The order of magnitude of the branching ratio shows a possibility to study such channels at the LHC.
\item The $\Lambda_{b}\rightarrow \Lambda \ell^{+}\ell^{-}$ transition is more probable, specially for $\tau$ case, in full theory in comparison with HQET.
\end{itemize}
\subsection{Forward Backward Asymmetry}
The lepton forward-backward asymmetry is one of the promising  tools in looking
for new physics beyond the SM such as extra dimensions.
This asymmetry is defined as:
 \bea {\cal A}_{FB} = \frac{N_f-N_b}{N_f+N_b}
 \eea where $N_f$ is the number of events that particle is moving "forward" with respect to
 any chosen direction, while $N_b$ is the number of events for
particle motion in "backward" direction.
 The
forward--backward asymmetry ${\cal A}_{FB}(\hat s,1/R)$ is defined in terms of the  differential 
decay rate as: \bea {\cal A}_{FB} (\hat s,1/R)=
\frac{\ds{\int_0^1\frac{d\Gamma}{d\hat{s}dz}}(z,\hat s,1/R)\,dz -
\ds{\int_{-1}^0\frac{d\Gamma}{d\hat{s}dz}}(z,\hat s,1/R)\,dz}
{\ds{\int_0^1\frac{d\Gamma}{d\hat{s}dz}}(z,\hat s,1/R)\,dz +
\ds{\int_{-1}^0\frac{d\Gamma}{d\hat{s}dz}}(z,\hat s,1/R)\,dz}~. \eea 
We depict the dependence of  ${\cal A}_{FB}(\hat s,1/R)$ asymmetry on $1/R$ for different leptons and at a fixed value of $\hat s=0.5$ common for allowed physical regions of all leptons in Fig. 2. 
\begin{figure}
\centering
\begin{tabular}{ccc}
\epsfig{file=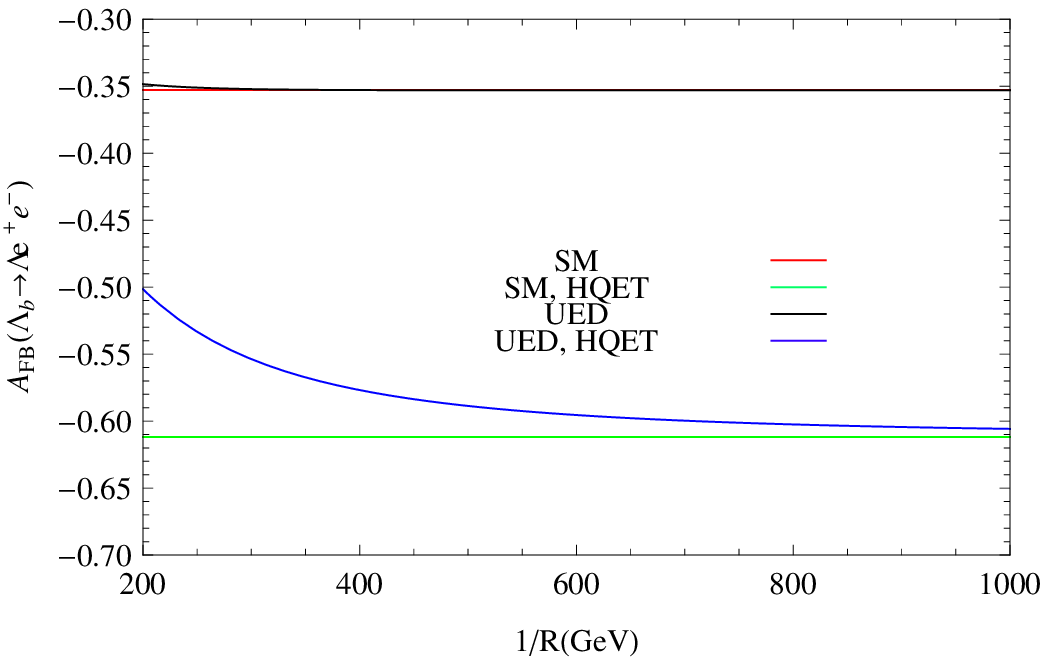,width=0.33\linewidth,clip=} &
\epsfig{file=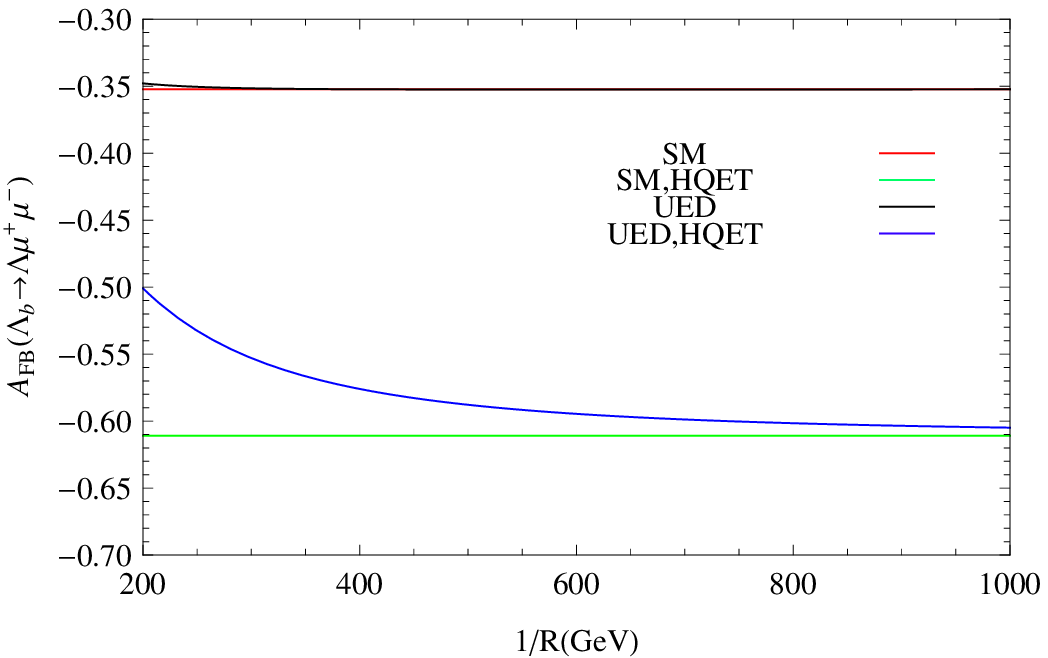,width=0.33\linewidth,clip=} &
\epsfig{file=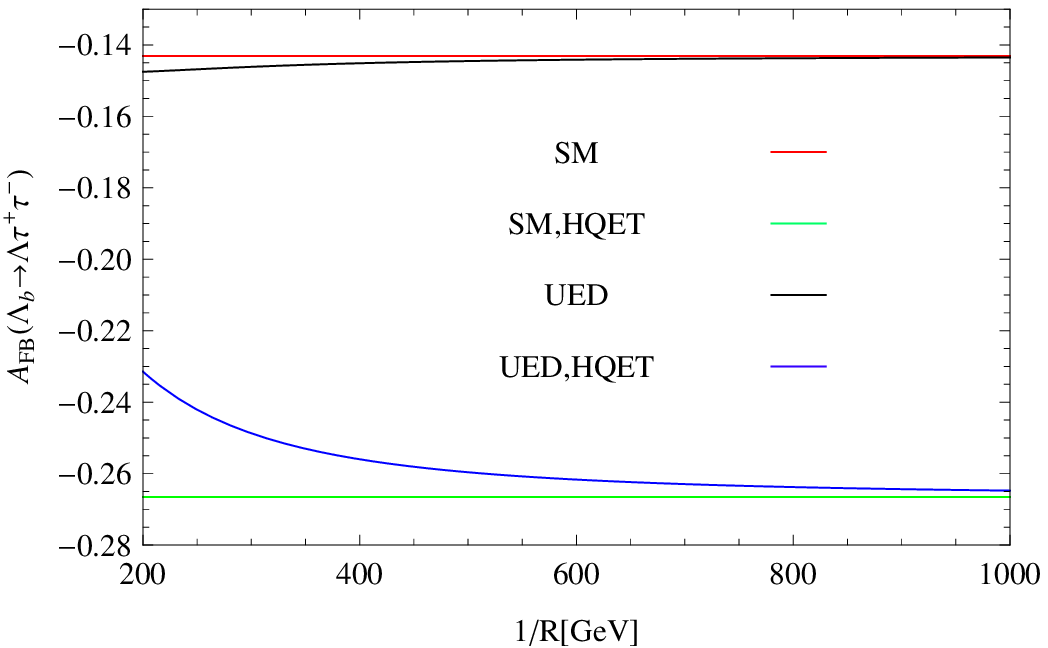,width=0.33\linewidth,clip=}
\end{tabular}
\caption{The dependence of ${\cal A}_{FB}(\hat s,1/R)$ asymmetry on compactification factor, $1/R$ for different leptons at $\hat s=0.5$.}
\end{figure}
A quick glance at these figures leads to the following results:
\begin{itemize}
\item  The ${\cal A}_{FB}$ is approximately the same for $e$ and $\mu$ and about 2-2.5 times greater than that of $\tau$ case.
\item As far as HQET is considered, there is  considerable discrepancy between the predictions of the ACD and SM models for  low  values of  $1/R$. As  $1/R$ increases, this difference starts to  diminish 
and at $1/R\simeq1000~GeV$, the two models have approximately the same results. In full theory, two models have approximately the same predictions for all leptons and all $1/R$ values.
\item For all leptons, the  forward-backward asymmetries show considerable differences between  the full theory and HQET predictions.
\end{itemize}
\subsection{ $\Lambda$ Baryon Polarizations} 
The definitions for polarizations of  $\Lambda$ baryon in  $\Lambda_b \rar \Lambda \ell^+ \ell^-$ channel are given in \cite{R7615}. Using those definitions, the $1/R$-dependent normal ($P_N$), transversal
  ($P_T$) and longitudinal 
($P_L$)  polarizations of the $\Lambda$ baryon in the massive lepton case are found as (for the general model independent case see \cite{30ozpineci,ozpinecibey}):
\bea \label{a2} P_N (\hat s, 1/R)\es \frac{8 \pi m_{\Lambda_b}^3 v
\sqrt{\hat s}}{\Delta (\hat s, 1/R)}\Bigg\{ 
- 2 m_{\Lambda_b} (1-r+\hat s) \sqrt{r} \,
\mbox{\rm Re}[A_1^\ast D_1 + B_1^\ast E_1] \nnb \\
\ar m_{\Lambda_b} (1-\sqrt{r}) [(1+\sqrt{r})^2 -\hat s] \, \Big( 
 m_\ell \mbox{\rm Re}[(A_2-B_2)^\ast F_1] \Big) \nnb \\
\ar m_\ell [(1+\sqrt{r})^2 -\hat s] \,
\mbox{\rm Re}[A_1^\ast F_1] \nnb \\
\ar 4 m_{\Lambda_b}^2 \hat s \sqrt{r} \, \mbox{\rm Re}[A_1^\ast E_2 +
A_2^\ast E_1 +B_1^\ast D_2 +
B_2^\ast D_1] \nnb \\
\ek 2 m_{\Lambda_b}^3 \hat s \sqrt{r} (1-r+\hat s) \,
\mbox{\rm Re}[A_2^\ast D_2 + B_2^\ast E_2^\ast] \nnb \\
\ar 2 m_{\Lambda_b} (1-r-\hat s) \, \Big( \mbox{\rm Re}[A_1^\ast E_1 +
B_1^\ast D_1] +
m_{\Lambda_b}^2 \hat s \mbox{\rm Re}[A_2^\ast E_2 + B_2^\ast D_2] \Big) \nnb \\
\ek m_{\Lambda_b}^2 [(1-r)^2-\hat s^2] \, \mbox{\rm Re}[A_1^\ast D_2 +
A_2^\ast D_1 + B_1^\ast E_2 +
B_2^\ast E_1] \nnb \\
\ek m_\ell [(1+\sqrt{r})^2 -\hat s] \,
\mbox{\rm Re}[B_1^\ast F_1] 
\Bigg\}~ , \eea
 \bea
\label{e12}
\lefteqn{
P_T(\hat s, 1/R) = - \frac{8 \pi m_{\Lambda_b}^3 v \sqrt{\hat s\lambda}}
{\Delta(\hat s, 1/R)}
\Bigg\{
m_\ell \Big(
\mbox{\rm Im}[(A_1+B_1)^\ast F_1]     
\Big)} \nnb \\
\ek m_\ell m_{\Lambda_b} \Big[
(1+\sqrt{r}) \, \mbox{\rm Im}[(A_2+B_2)^\ast F_1] \Big] \nnb \\
\ar m_{\Lambda_b}^2 (1-r+\hat s) \Big(
\mbox{\rm Im}[A_2^\ast D_1 - A_1^\ast D_2] -
\mbox{\rm Im}[B_2^\ast E_1 - B_1^\ast E_2] \Big)\nnb \\
\ar 2 m_{\Lambda_b} \Big(
\mbox{\rm Im}[A_1^\ast E_1-B_1^\ast D_1] -
m_{\Lambda_b}^2 \hat s \, \mbox{\rm Im}[A_2^\ast E_2 - B_2^\ast D_2] \Big)\Bigg\}~,
\eea
\bea \label{a1} P_L (\hat s, 1/R)\es
\frac{16 m_{\Lambda_b}^2 \sqrt{\lambda}}{\Delta(\hat s, 1/R)} \Bigg\{ 8 m_\ell^2
m_{\Lambda_b}\, \Big( \mbox{\rm Re}[D_1^\ast E_3 - D_3^\ast E_1] +
\sqrt{r} \mbox{\rm Re}[D_1^\ast D_3 - E_1^\ast E_3)] \Big) \nnb \\
\ar 2 m_\ell m_{\Lambda_b}\,  (1+\sqrt{r}) \mbox{\rm
Re}[(D_1-E_1)^\ast F_2]
 \nnb \\
\ek 2 m_\ell m_{\Lambda_b}^2 \hat s \, \Big\{ \mbox{\rm
Re}[(D_3-E_3)^\ast F_2 ] +
2 m_\ell ( \vel D_3 \ver^2 - \vel E_3 \ver^2 ) \Big\} \nnb \\
\ek 4 m_{\Lambda_b} (2 m_\ell^2 + m_{\Lambda_b}^2 \hat s) \,
\mbox{\rm Re}[A_1^\ast B_2 - A_2^\ast B_1] \nnb \\
\ek \frac{4}{3} m_{\Lambda_b}^3 \hat s v^2 \, \Big( 3 \mbox{\rm
Re}[D_1^\ast E_2 - D_2^\ast E_1] +
\sqrt{r} \mbox{\rm Re}[D_1^\ast D_2 - E_1^\ast E_2] \Big) \nnb \\
\ek \frac{4}{3} m_{\Lambda_b} \sqrt{r} (6 m_\ell^2 +
m_{\Lambda_b}^2 \hat s v^2) \, \mbox{\rm Re}[A_1^\ast A_2 - B_1^\ast B_2] \nnb \\
\ar \frac{1}{3} \Big\{ 3 [4 m_\ell^2 + m_{\Lambda_b}^2 (1-r+\hat s)]
(\vel A_1 \ver^2 -
\vel B_1 \ver^2 ) - 3 [4 m_\ell^2 -  m_{\Lambda_b}^2 (1-r+\hat s)] \nnb \\
\cp (\vel D_1 \ver^2 - \vel E_1 \ver^2 ) -  m_{\Lambda_b}^2
(1-r-\hat s) v^2 (\vel A_1 \ver^2 - \vel B_1 \ver^2 + \vel D_1 \ver^2 -
\vel E_1 \ver^2 )
\Big\} \nnb \\
\ek \frac{1}{3} m_{\Lambda_b}^2 \{ 12 m_\ell^2 (1-r) +
m_{\Lambda_b}^2 \hat s [3 (1-r+\hat s) + v^2 (1-r-\hat s)] \}
(\vel A_2 \ver^2 - \vel B_2 \ver^2) \nnb \\
\ek \frac{2}{3} m_{\Lambda_b}^4 \hat s (2 - 2 r + \hat s) v^2 \,
(\vel D_2 \ver^2 - \vel E_2 \ver^2) 
\Bigg\}~,  \eea 
 where,
 \bea \Delta(\hat s, 1/R)\es{{\cal T}_0(\hat s, 1/R) +\frac{1}{3} {\cal T}_2(\hat s, 1/R)}.
 \eea
For instance, we show the dependence of  the  $P_N$ and $P_T$ polarizations of the $\Lambda$ baryon on compactification factor at a fixed value of $\hat s=0.5$ in Figs. 3 and 4 , respectively. 
From these figures, we infer the following information:
\begin{itemize}
 \item In the case of $P_N$ and  all leptons, we observe a (25-35)\% HQET violations. This violation is very small for the transverse polarization of the $\Lambda$.
\item The UED predictions deviate considerably from the  SM results in the case of $P_T$ and small values of the compactification factor. This deviation is  small for the   $P_N$ compared to the $P_T$.
 In the case of $\tau$ and HQET, two models have approximately the same predictions for the normal polarization.
\end{itemize}

%\begin{figure}
%\centering
%\begin{tabular}{ccc}
%\epsfig{file=leR.eps,width=0.33\linewidth,clip=} &
%\epsfig{file=lmuonR.eps,width=0.33\linewidth,clip=} &
%\epsfig{file=ltauR.eps,width=0.33\linewidth,clip=}
%\end{tabular}
%\caption{The dependence of $P_L(\hat s,1/R)$  on compactification factor, $1/R$ for different leptons at $\hat s=0.5$.}
%\end{figure}
\begin{figure}
\centering
\begin{tabular}{ccc}
\epsfig{file=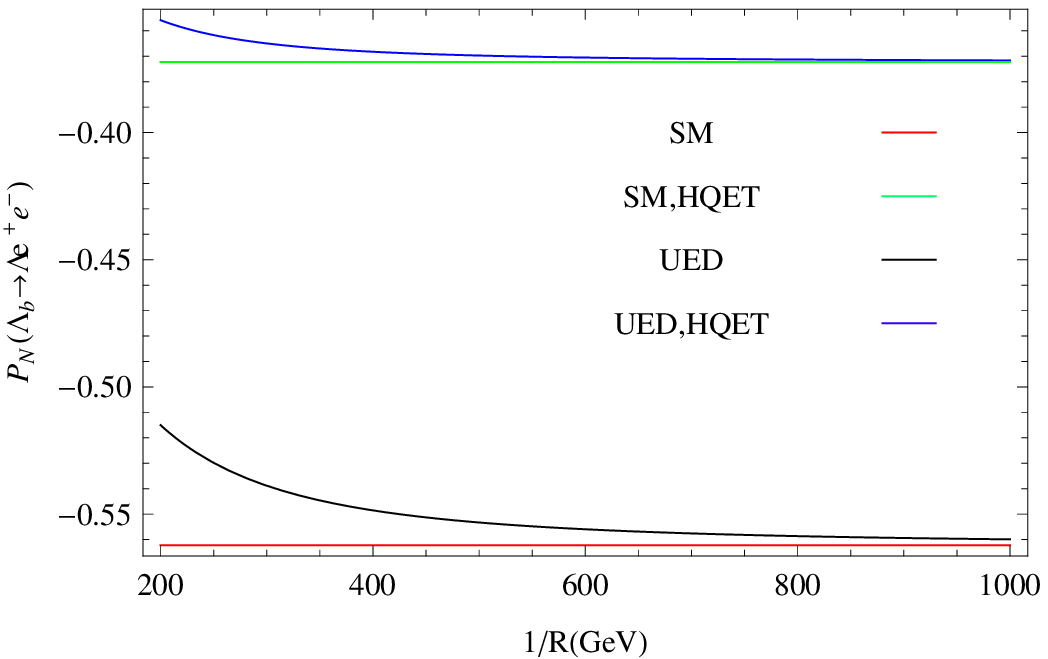,width=0.33\linewidth,clip=} &
\epsfig{file=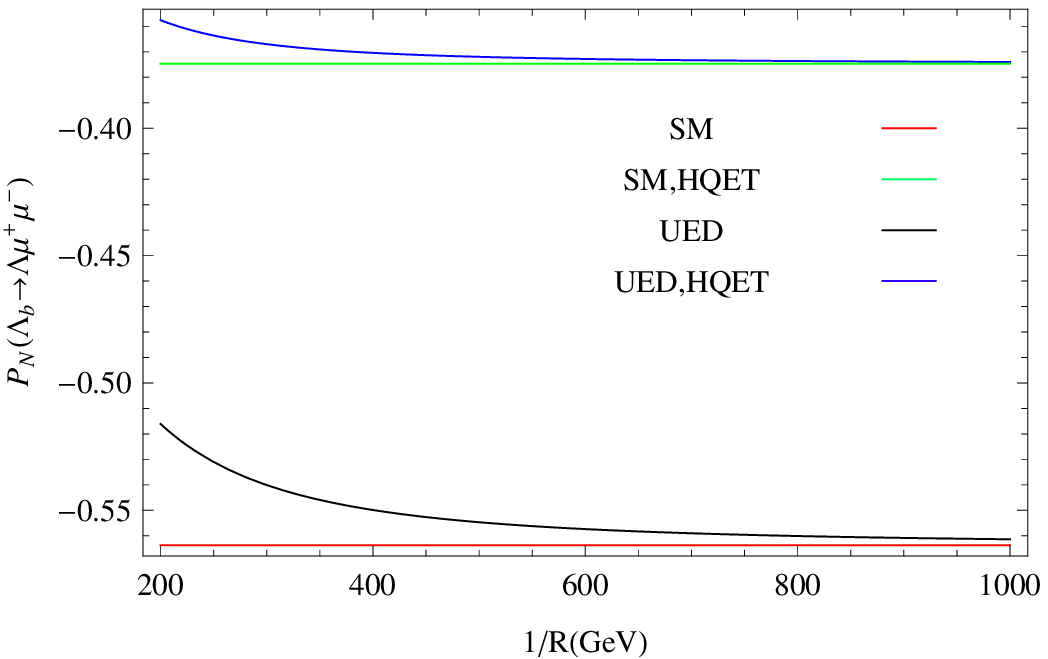,width=0.33\linewidth,clip=} &
\epsfig{file=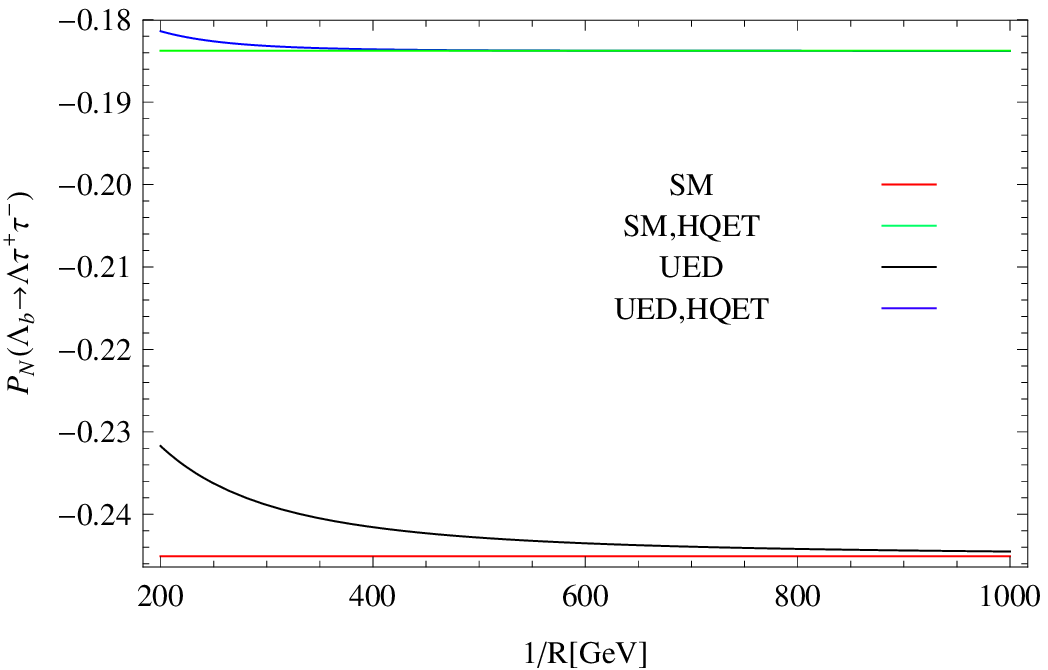,width=0.33\linewidth,clip=}
\end{tabular}
\caption{The dependence of $P_N(\hat s,1/R)$  on compactification factor, $1/R$ for different leptons at $\hat s=0.5$.}
\end{figure}
\begin{figure}
\centering
\begin{tabular}{ccc}
\epsfig{file=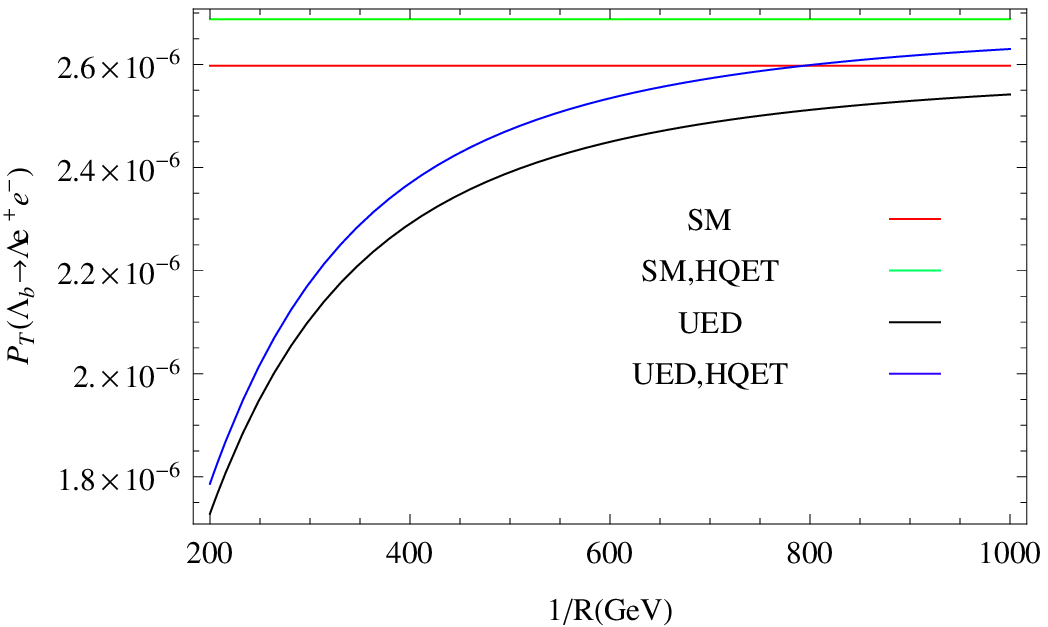,width=0.33\linewidth,clip=} &
\epsfig{file=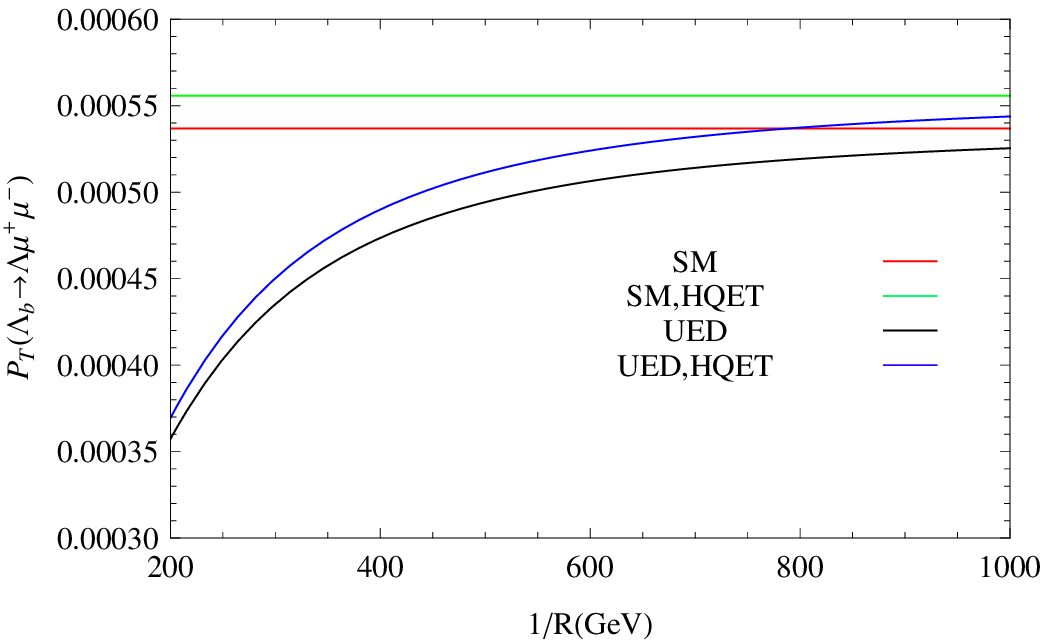,width=0.33\linewidth,clip=} &
\epsfig{file=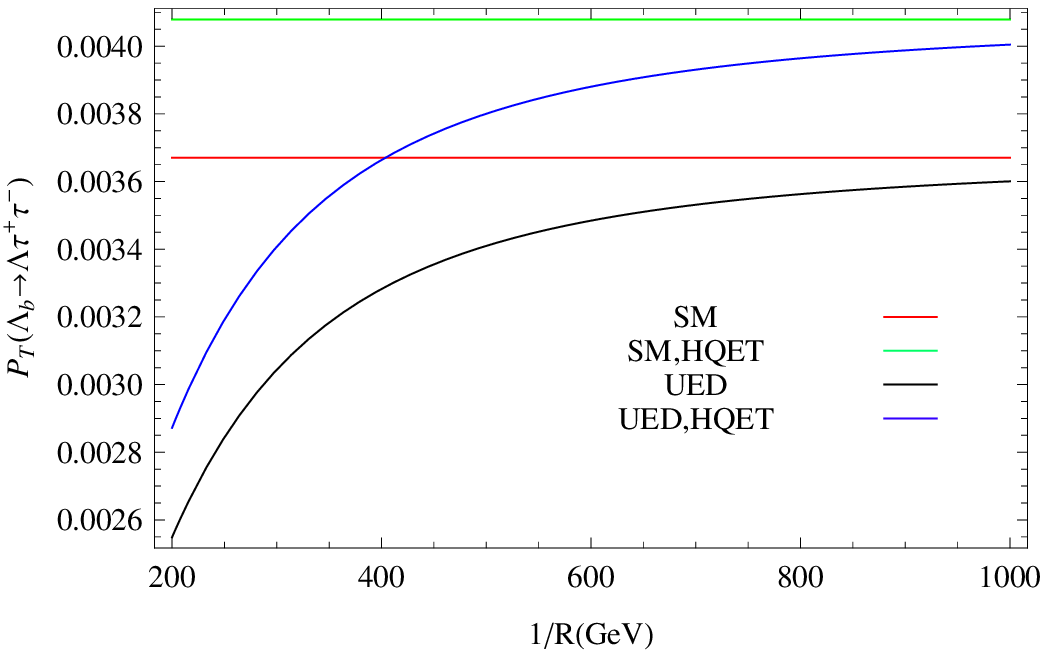,width=0.33\linewidth,clip=}
\end{tabular}
\caption{The dependence of $P_T(\hat s,1/R)$  on compactification factor, $1/R$ for different leptons at $\hat s=0.5$.}
\end{figure}
\subsection{ Double Lepton Polarization Asymmetries}
 The double lepton polarization asymmetries related to the $\Lambda_b \rar \Lambda \ell^+ \ell^-$ transition are defined in \cite{vahli} for general model independent form of the effective Hamiltonian.
 In our case, in the rest frame of
$\ell^\pm$,  the $1/R$-dependent double longitudinal, transverse and normal asymmetries are obtained as (see also \cite{0608143,R7710}):

%\begin{figure}
%\centering
%\begin{tabular}{ccc}
%\epsfig{file=lleR.eps,width=0.33\linewidth,clip=} &
%\epsfig{file=llmuonR.eps,width=0.33\linewidth,clip=} &
%\epsfig{file=lltauR.eps,width=0.33\linewidth,clip=}
%\end{tabular}
%\caption{The dependence of $P_{LL}(\hat s,1/R)$  on compactification factor, $1/R$ for different leptons at $\hat s=0.5$.}
%\end{figure}
\bea
 \label{e7717} P_{LN}(\hat s, 1/R) \es \frac{16 \pi m_{\Lambda_b}^4
\hat{m}_\ell \sqrt{\lambda}}{\Delta(\hat s, 1/R) \sqrt{\hat{s}}} \mbox{\rm Im}
\Bigg\{
%1
(1-r) (A_1^\ast D_1 + B_1^\ast E_1)
%2
%3
%4
+ m_{\Lambda_b}
 \hat{s} (A_1^\ast E_3 - A_2^\ast E_1 + B_1^\ast D_3
-B_2^\ast D_1) \nnb \\
%5
%6
%7
%8
%9
\ar m_{\Lambda_b}
 \sqrt{r} \hat{s}
(A_1^\ast D_3 + A_2^\ast D_1 +B_1^\ast E_3 + B_2^\ast E_1)
%10
%11
- m_{\Lambda_b}^2 \hat{s}^2 \Big( B_2^\ast E_3 + A_2^\ast D_3
\Big) \Bigg\},\eea
\begin{figure}
\centering
\begin{tabular}{ccc}
\epsfig{file=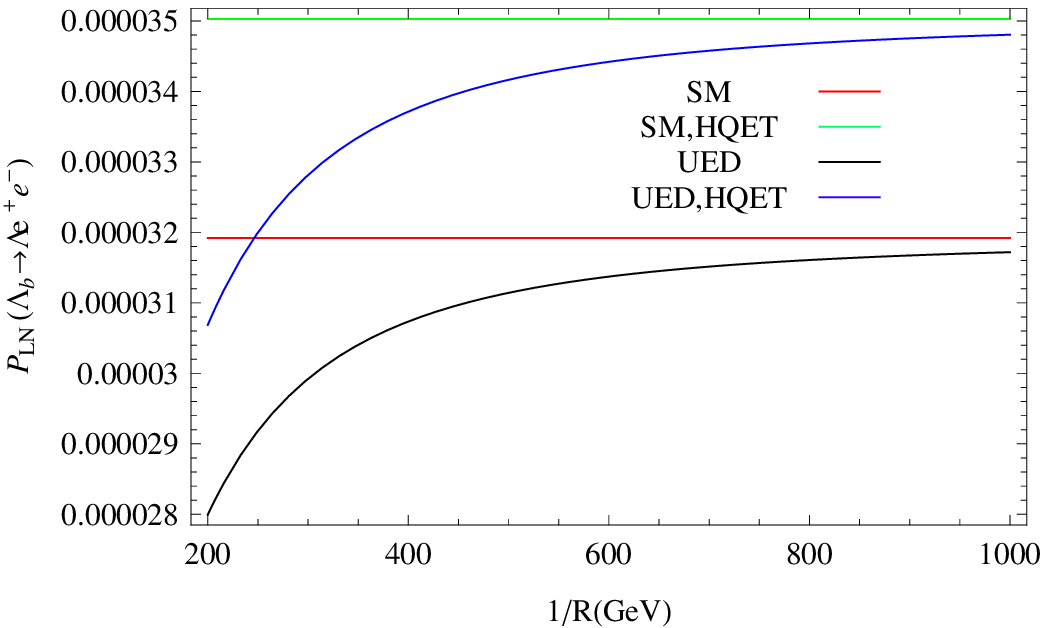,width=0.33\linewidth,clip=} &
\epsfig{file=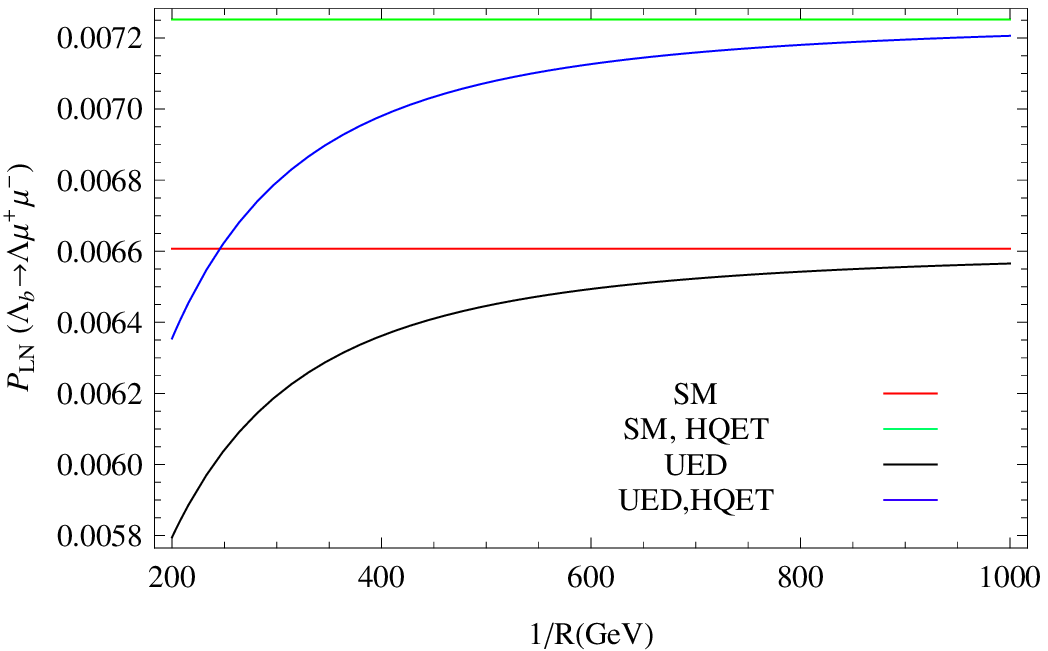,width=0.33\linewidth,clip=} &
\epsfig{file=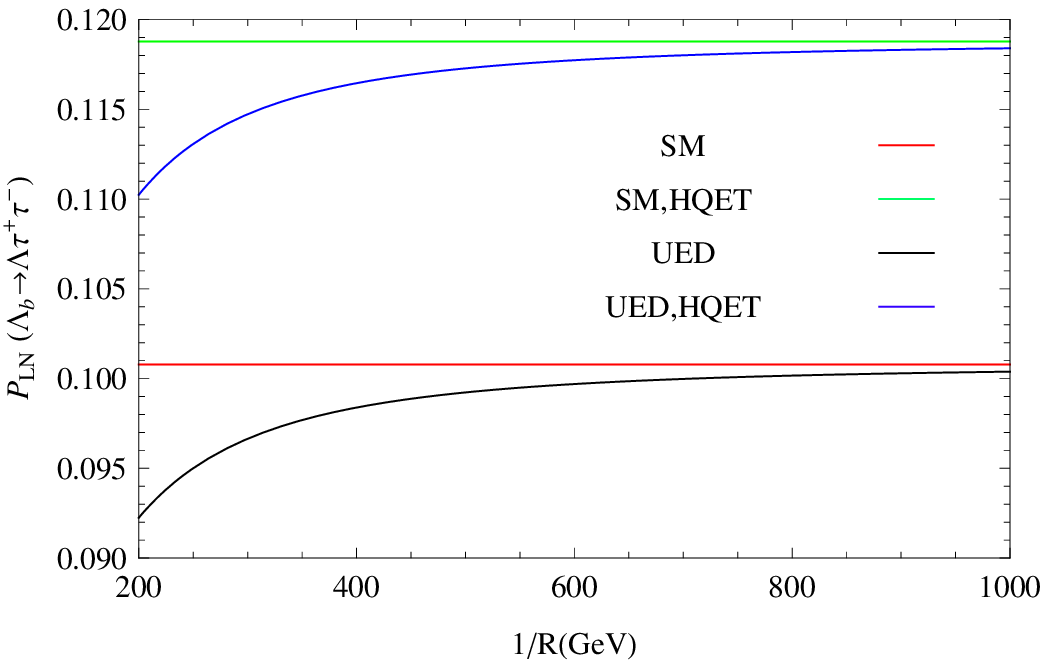,width=0.33\linewidth,clip=}
\end{tabular}
\caption{The dependence of $P_{LN}(\hat s,1/R)$  on compactification factor, $1/R$ for different leptons at $\hat s=0.5$.}
\end{figure}
\bea
 \label{e7718} P_{NL} (\hat s, 1/R)\es - \frac{16 \pi m_{\Lambda_b}^4
\hat{m}_\ell \sqrt{\lambda}}{\Delta \sqrt{\hat{s}}} \mbox{\rm Im}
\Bigg\{
%1
(1-\hat{r}_\Lambda) (A_1^\ast D_1 + B_1^\ast E_1)
%2
%3
%4
+ m_{\Lambda_b}
 \hat{s} (A_1^\ast E_3 - A_2^\ast E_1 + B_1^\ast D_3
-B_2^\ast D_1) \nnb \\
%5
%6
%7
%8
%9
\ek m_{\Lambda_b}
 \sqrt{\hat{r}_\Lambda} \hat{s}
(A_1^\ast D_3 + A_2^\ast D_1 +B_1^\ast E_3 + B_2^\ast E_1)
%10
%11
- m_{\Lambda_b}^2 \hat{s}^2 \Big( B_2^\ast E_3 + A_2^\ast D_3
\Big) \Bigg\},\eea

\bea \label{e7719} P_{LT}(\hat s, 1/R) \es \frac{16 \pi m_{\Lambda_b}^4
\hat{m}_\ell \sqrt{\lambda} v}{\Delta(\hat s, 1/R) \sqrt{\hat{s}}} \mbox{\rm
Re} \Bigg\{
%1
(1-r) \Big( \vel D_1 \ver^2 + \vel E_1 \ver^2 \Big)
%2
- \hat{s} \Big(A_1 D_1^\ast - B_1 E_1^\ast \Big) \nnb \\
%3
\ek m_{\Lambda_b} \hat{s} \Big[ B_1 D_2^\ast + (A_2 + D_2 -D_3)
E_1^\ast -  A_1 E_2^\ast
-(B_2-E_2+E_3) D_1^\ast \Big] \nnb \\
%4
%5
%6
%8
\ar m_{\Lambda_b}
 \sqrt{r} \hat{s}
\Big[ A_1 D_2^\ast + (A_2 + D_2 +D_3) D_1^\ast - B_1 E_2^\ast -
(B_2 - E_2 - E_3) E_1^\ast \Big] \nnb \\
%7
\ar m_{\Lambda_b}^2 \hat{s} (1-r) (A_2 D_2^\ast -
B_2 E_2^\ast)
%9
- m_{\Lambda_b}^2 \hat{s}^2 (D_2 D_3^\ast + E_2 E_3^\ast )\Bigg\}, \eea
\begin{figure}
\centering
\begin{tabular}{ccc}
\epsfig{file=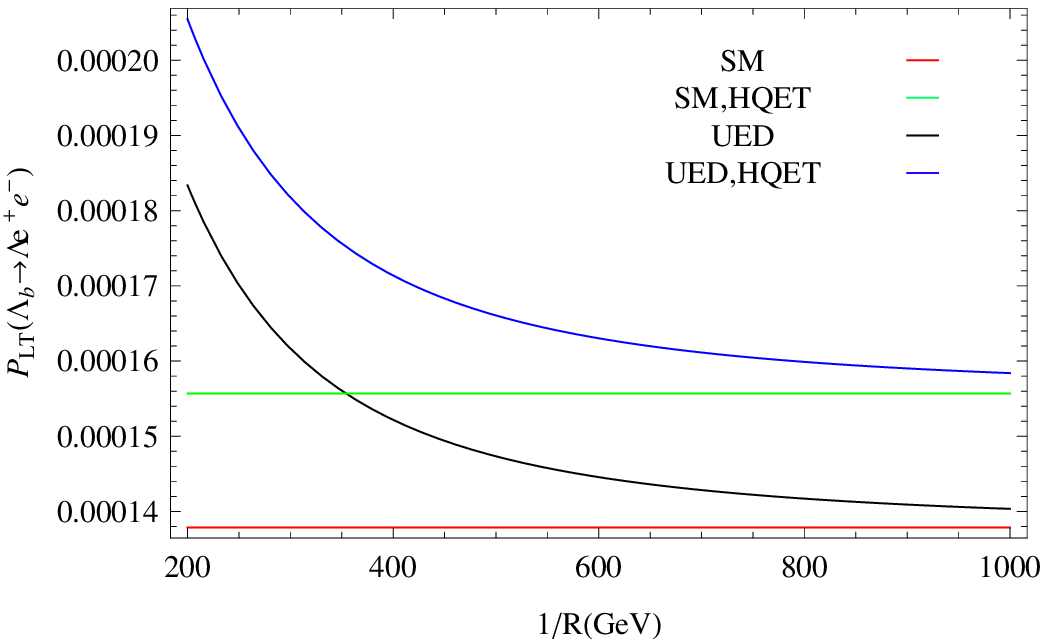,width=0.33\linewidth,clip=} &
\epsfig{file=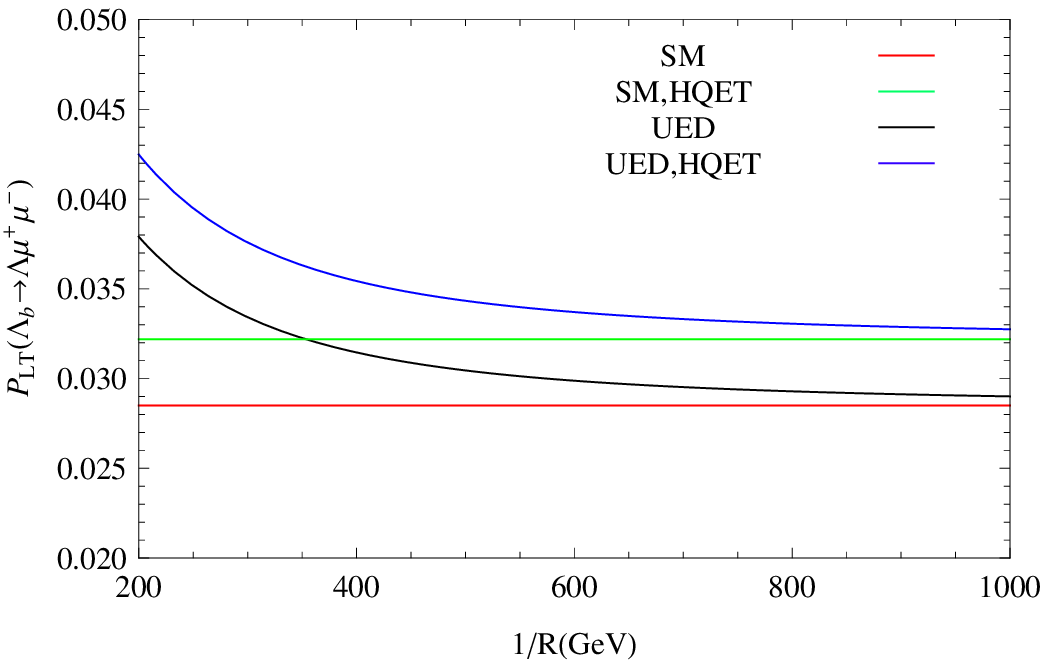,width=0.33\linewidth,clip=} &
\epsfig{file=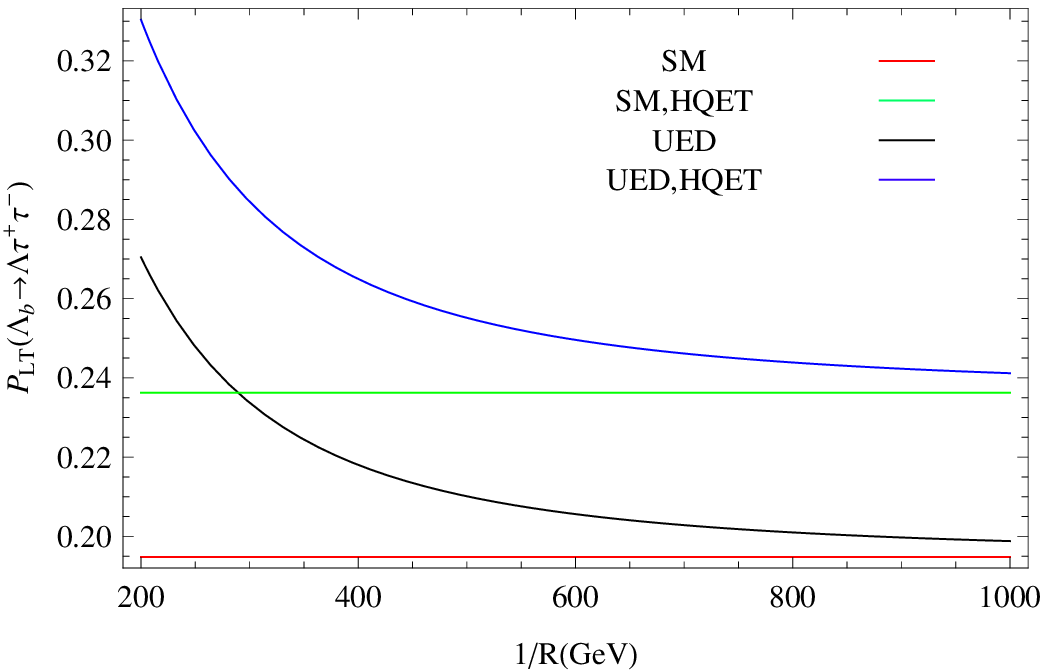,width=0.33\linewidth,clip=}
\end{tabular}
\caption{The dependence of $P_{LT}(\hat s,1/R)$  on compactification factor, $1/R$ for different leptons at $\hat s=0.5$.}
\end{figure}
\bea \label{e7720} P_{TL}(\hat s, 1/R) \es \frac{16 \pi m_{\Lambda_b}^4
\hat{m}_\ell \sqrt{\lambda} v}{\Delta \sqrt{\hat{s}}} \mbox{\rm
Re} \Bigg\{
%1
(1-\hat{r}_\Lambda) \Big( \vel D_1 \ver^2 + \vel E_1 \ver^2 \Big)
%2
+ \hat{s} \Big(A_1 D_1^\ast - B_1 E_1^\ast \Big) \nnb \\
%3
\ar m_{\Lambda_b} \hat{s} \Big[ B_1 D_2^\ast + (A_2 - D_2 + D_3)
E_1^\ast -  A_1 E_2^\ast
- (B_2+E_2-E_3) D_1^\ast \Big] \nnb \\
%4
%5
%6
%8
\ek m_{\Lambda_b}
 \sqrt{\hat{r}_\Lambda} \hat{s}
\Big[ A_1 D_2^\ast + (A_2 - D_2 - D_3) D_1^\ast - B_1 E_2^\ast -
(B_2 + E_2 + E_3) E_1^\ast \Big] \nnb \\
%7
\ek m_{\Lambda_b}^2 \hat{s} (1-\hat{r}_\Lambda) (A_2 D_2^\ast -
B_2 E_2^\ast)
%9
- m_{\Lambda_b}^2 \hat{s}^2 (D_2 D_3^\ast + E_2 E_3^\ast ) \Bigg\},\eea
\bea \label{e7716} P_{LL}(\hat s, 1/R) \es \frac{16 m_{\Lambda_b}^4}{3\Delta(\hat s, 1/R)}
\mbox{\rm Re} \Bigg\{\nnb \\
%1
%2
%3
%4
%5
%6
\ek 6 m_{\Lambda_b} \sqrt{r}
(1-r+\hat{s}) \Big[ \hat{s} (1+v^2) (A_1 A_2^\ast +
B_1 B_2^\ast)  -
4 \hat{m}_\ell^2 (D_1 D_3^\ast + E_1 E_3^\ast) \Big] \nnb \\
%7
%8
\ar 6 m_{\Lambda_b} (1-r-\hat{s}) \Big[ \hat{s}
(1+v^2) (A_1 B_2^\ast + A_2 B_1^\ast) +
4 \hat{m}_\ell^2 (D_1 E_3^\ast + D_3 E_1^\ast) \Big] \nnb \\
%9
\ar 12 \sqrt{r} \hat{s} (1+v^2) \Big( A_1 B_1^\ast +
D_1 E_1^\ast +
m_{\Lambda_b}^2 \hat{s} A_2 B_2^\ast \Big) \nnb \\
%10
\ar 12 m_{\Lambda_b}^2 \hat{m}_\ell^2 \hat{s}
(1+r-\hat{s})
\ga \vel D_3 \ver^2 + \vel E_3^\ast \ver^2 \dr \nnb \\
%11
%12
\ek (1+v^2) \Big[ 1+r^2 - r
(2-\hat{s}) +\hat{s} (1-2 \hat{s}) \Big]
\Big(\vel A_1 \ver^2 + \vel B_1 \ver^2 \Big) \nnb \\
%13
\ek \Big[ (5 v^2 - 3) (1-r)^2 + 4 \hat{m}_\ell^2
(1+r) + 2 \hat{s} (1+8 \hat{m}_\ell^2 +
r)
- 4 \hat{s}^2 \Big] \Big( \vel D_1 \ver^2 + \vel E_1 \ver^2 \Big) \nnb \\
%14
\ek m_{\Lambda_b}^2 (1+v^2) \hat{s} \Big[2 + 2 r^2
-\hat{s}(1 +\hat{s}) - r (4 + \hat{s})\Big] \big(
\vel A_2 \ver^2 + \vel B_2 \ver^2 \Big) \nnb \\
%15
%16
\ek 2 m_{\Lambda_b}^2 \hat{s} v^2 \Big[ 2 (1 + r^2)
- \hat{s} (1+\hat{s}) - r (4+\hat{s})\Big] \Big(
\vel D_2 \ver^2 + \vel E_2 \ver^2 \Big) \nnb \\
%17
\ar 12 m_{\Lambda_b} \hat{s} (1-r-\hat{s}) v^2
\Big( D_1 E_2^\ast + D_2 E_1^\ast \Big) \nnb \\
%18
\ek 12 m_{\Lambda_b} \sqrt{r} \hat{s}
(1-r+\hat{s}) v^2
\Big( D_1 D_2^\ast + E_1 E_2^\ast \Big) \nnb \\
%19
\ar 24 m_{\Lambda_b}^2 \sqrt{r} \hat{s} \Big(
\hat{s} v^2 D_2 E_2^\ast + 2 \hat{m}_\ell^2 D_3 E_3^\ast \Big)\Bigg\},
\eea

\bea \label{e7721} P_{NT} (\hat s, 1/R)\es \frac{64 m_{\Lambda_b}^4 \lambda
v}{3 \Delta(\hat s, 1/R)} \mbox{\rm Im} \Bigg\{
%1
(A_1 D_1^\ast +B_1 E_1^\ast)
%2
%3
%4
%5
%6
%7
+ m_{\Lambda_b}^2 \hat{s} (A_2^\ast D_2 + B_2^\ast E_2)\Bigg\},
%8
%9
\eea
\begin{figure}
\centering
\begin{tabular}{ccc}
\epsfig{file=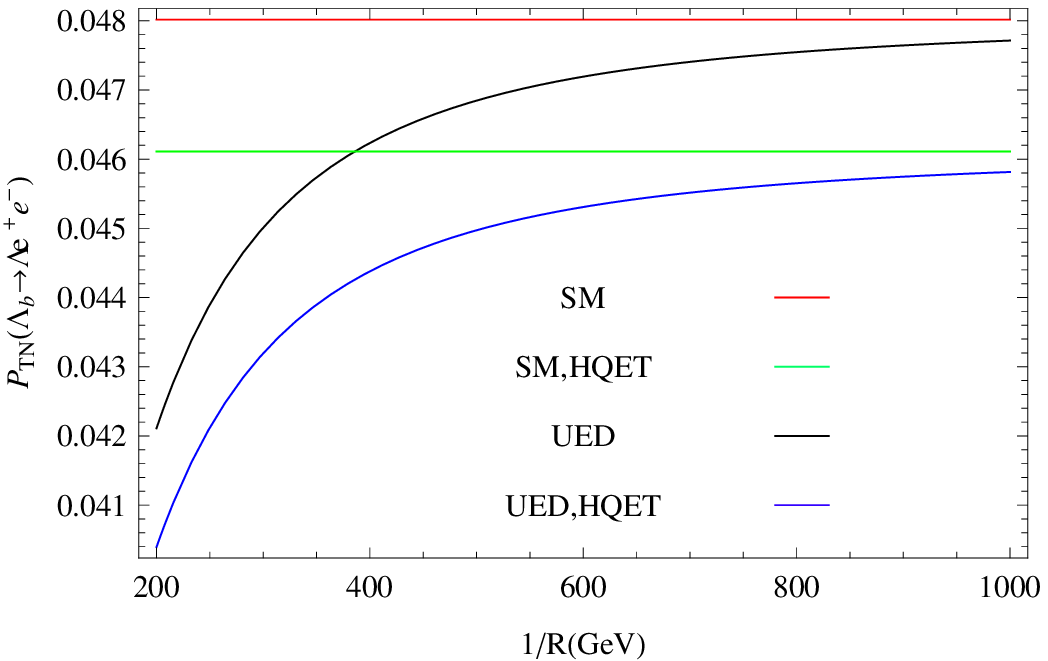,width=0.33\linewidth,clip=} &
\epsfig{file=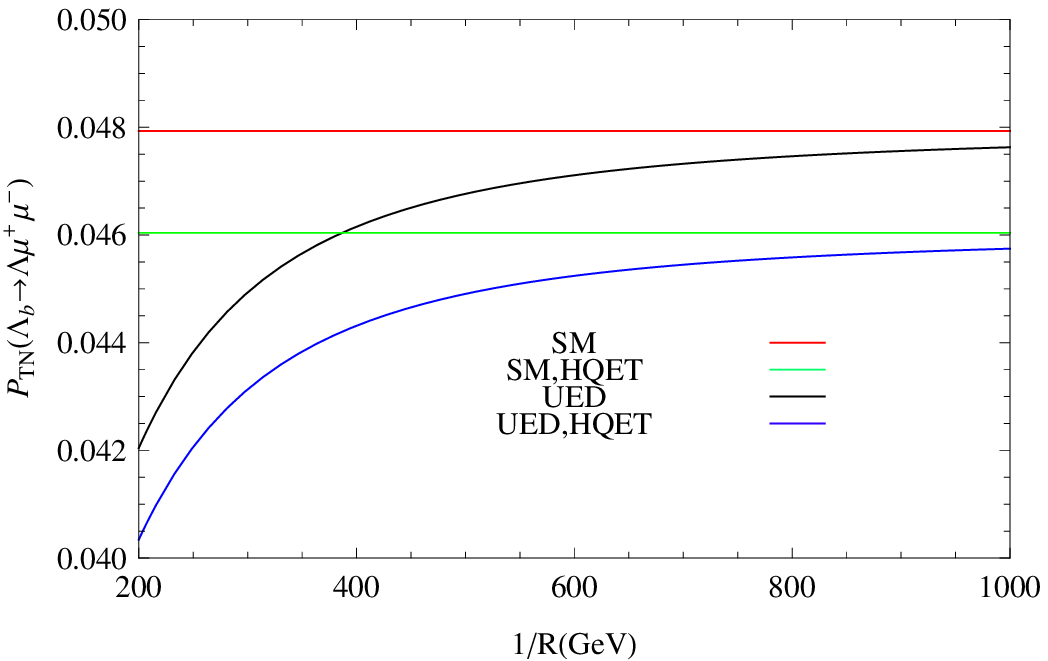,width=0.33\linewidth,clip=} &
\epsfig{file=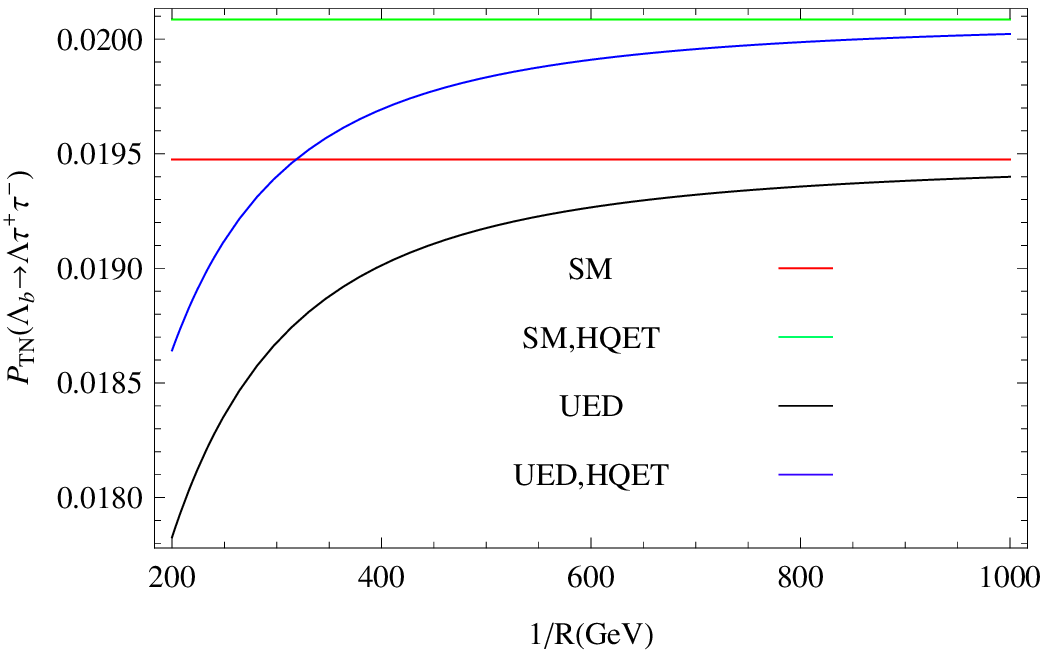,width=0.33\linewidth,clip=}
\end{tabular}
\caption{The dependence of $P_{TN}(\hat s,1/R)$  on compactification factor, $1/R$ for different leptons at $\hat s=0.5$.}
\end{figure}

\bea \label{e7722} P_{TN}(\hat s, 1/R) \es - \frac{64 m_{\Lambda_b}^4 \lambda
v}{3 \Delta} \mbox{\rm Im} \Bigg\{
%1
(A_1 D_1^\ast +B_1 E_1^\ast)
%2
%3
%4
%5
%6
%7
+ m_{\Lambda_b}^2 \hat{s} (A_2^\ast D_2 + B_2^\ast E_2)\Bigg\},
%8
%9
\eea

\bea \label{e7723} P_{NN} (\hat s, 1/R)\es \frac{32 m_{\Lambda_b}^4}{3 \hat{s}
\Delta(\hat s, 1/R)} \mbox{\rm Re} \Bigg\{
%1
%2
24 \hat{m}_\ell^2 \sqrt{r} \hat{s}
( A_1 B_1^\ast + D_1 E_1^\ast ) \nnb \\
%3
\ek 12 m_{\Lambda_b} \hat{m}_\ell^2 \sqrt{r} \hat{s}
(1-r +\hat{s}) (A_1 A_2^\ast + B_1 B_2^\ast) \nnb \\
%4
%5
%6
%7
%8
%9
\ar 6 m_{\Lambda_b} \hat{m}_\ell^2 \hat{s} \Big[ m_{\Lambda_b}
\hat{s} (1+r-\hat{s}) \Big(\vel D_3 \ver^2 + \vel
E_3 \ver^2 \Big) + 2 \sqrt{r}
(1-r+\hat{s})
(D_1 D_3^\ast + E_1 E_3^\ast)\Big] \nnb \\
%10
\ar 12 m_{\Lambda_b} \hat{m}_\ell^2 \hat{s}
(1-r-\hat{s})
(A_1 B_2^\ast + A_2 B_1 ^\ast + D_1 E_3^\ast + D_3 E_1^\ast) \nnb \\
%11
\ek [ \lambda \hat{s} + 2 \hat{m}_\ell^2 (1 + r^2 -
2 r + r \hat{s} + \hat{s} - 2
\hat{s}^2) ] \Big( \vel A_1 \ver^2 + \vel B_1 \ver^2 - \vel D_1
\ver^2 -
\vel E_1 \ver^2 \Big) \nnb \\
%12
%13
\ar 24 m_{\Lambda_b}^2 \hat{m}_\ell^2 \sqrt{r}
\hat{s}^2 (A_2 B_2^\ast + D_3 E_3^\ast)
%14
%15
- m_{\Lambda_b}^2 \lambda \hat{s}^2 v^2
\Big( \vel D_2 \ver^2 + \vel E_2 \ver^2 \Big) \nnb \\
%16
\ar m_{\Lambda_b}^2 \hat{s} \{ \lambda \hat{s} - 2 \hat{m}_\ell^2
[2 (1+ r^2) - \hat{s} (1+\hat{s}) - r
(4+\hat{s})]\} \Big( \vel A_2 \ver^2 + \vel B_2 \ver^2 \Big)\Bigg\},
%17
%18
\eea
\begin{figure}
\centering
\begin{tabular}{ccc}
\epsfig{file=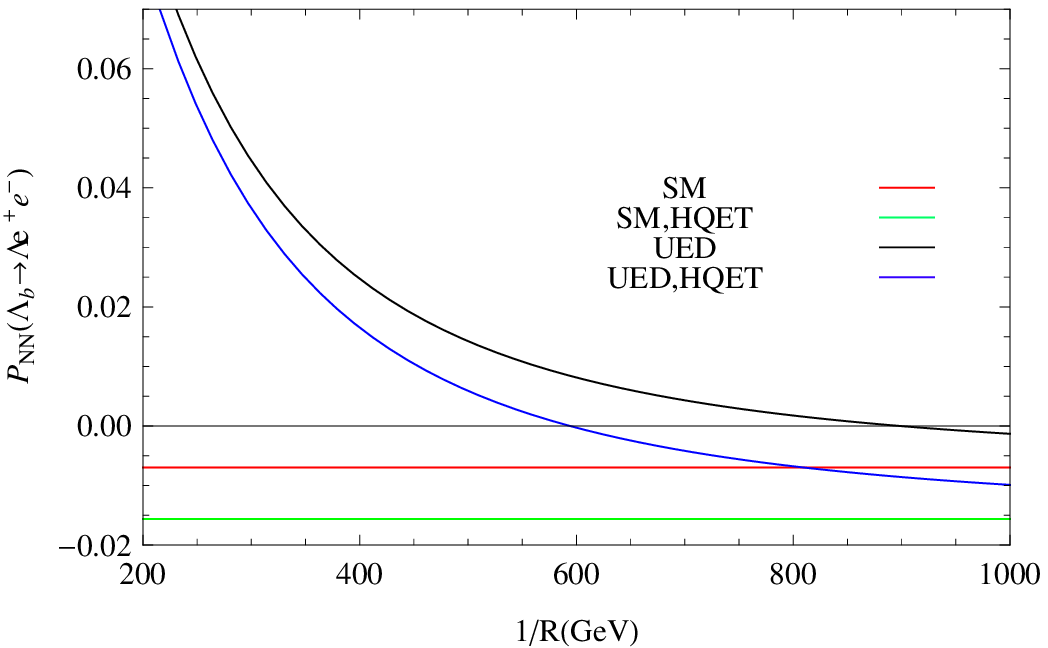,width=0.33\linewidth,clip=} &
\epsfig{file=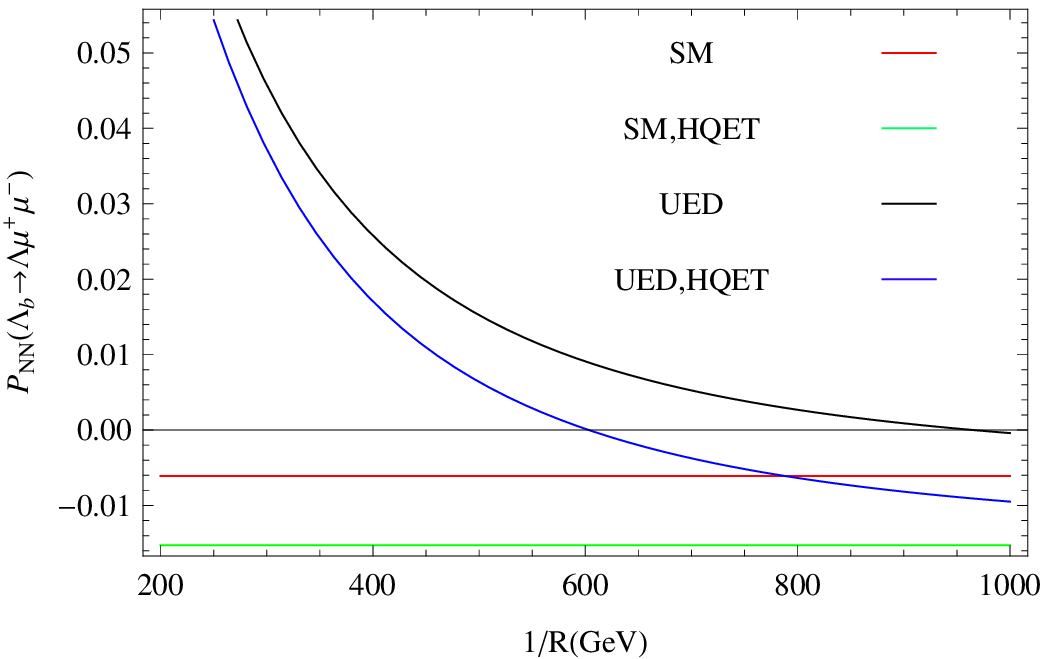,width=0.33\linewidth,clip=} &
\epsfig{file=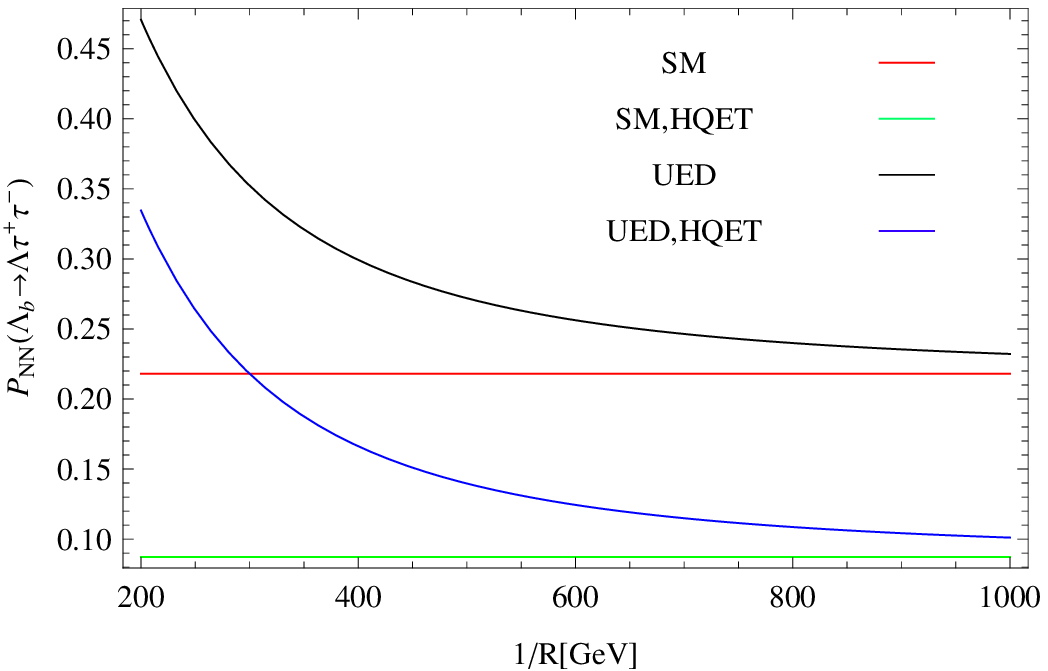,width=0.33\linewidth,clip=}
\end{tabular}
\caption{The dependence of $P_{NN}(\hat s,1/R)$  on compactification factor, $1/R$ for different leptons at $\hat s=0.5$.}
\end{figure}
\bea \label{e7724} P_{TT}(\hat s, 1/R) \es \frac{32 m_{\Lambda_b}^4}{3 \hat{s}
\Delta(\hat s, 1/R)} \mbox{\rm Re} \Bigg\{
%1
%2
- 24 \hat{m}_\ell^2 \sqrt{r} \hat{s}
( A_1 B_1^\ast + D_1 E_1^\ast ) \nnb \\
%3
\ek 12 m_{\Lambda_b} \hat{m}_\ell^2 \sqrt{r} \hat{s}
(1-r +\hat{s}) (D_1 D_3^\ast + E_1 E_3^\ast)
%4
%5
%6
- 24 m_{\Lambda_b}^2 \hat{m}_\ell^2 \sqrt{r}
\hat{s}^2
( A_2 B_2^\ast + D_3 E_3^\ast ) \nnb \\
%7
%8
%9
\ek 6 m_{\Lambda_b} \hat{m}_\ell^2 \hat{s} \Big[ m_{\Lambda_b}
\hat{s} (1+r-\hat{s}) \Big(\vel D_3 \ver^2 + \vel
E_3 \ver^2 \Big) - 2 \sqrt{r}
(1-r+\hat{s})
(A_1 A_2^\ast + B_1 B_2^\ast)\Big] \nnb \\
%10
\ek 12 m_{\Lambda_b} \hat{m}_\ell^2 \hat{s}
(1-r-\hat{s})
(A_1 B_2^\ast + A_2 B_1 ^\ast + D_1 E_3^\ast + D_3 E_1^\ast) \nnb \\
%11
\ek [ \lambda \hat{s} - 2 \hat{m}_\ell^2 (1 + r^2 -
2 r + r \hat{s} + \hat{s} - 2
\hat{s}^2) ]
\Big( \vel A_1 \ver^2 + \vel B_1 \ver^2 \Big) \nnb \\
%12
\ar m_{\Lambda_b}^2 \hat{s} \{ \lambda \hat{s} + \hat{m}_\ell^2 [4
(1- r)^2 - 2 \hat{s} (1+r) - 2
\hat{s}^2 ]\}
\Big( \vel A_2 \ver^2 + \vel B_2 \ver^2 \Big) \nnb \\
%13
\ar \{ \lambda \hat{s} - 2 \hat{m}_\ell^2 [5 (1-
r)^2 - 7 \hat{s} (1+r) + 2 \hat{s}^2
]\}
\Big( \vel D_1 \ver^2 + \vel E_1 \ver^2 \Big) \nnb \\
\ek m_{\Lambda_b}^2 \lambda \hat{s}^2 v^2 \Big( \vel D_2 \ver^2 +
\vel E_2 \ver^2 \Big) \Bigg\}, \eea
where, $\hat m_l=\frac{m_l}{m_{\Lambda_b}}$. As examples, we depict the $1/R$ dependence of some double lepton polarization asymmetries at a fixed value of $\hat s=0.5$  in Figs. 5-9. From these figures, we obtain the following
 conclusions:
\begin{itemize}
 \item In all cases, there are substantial differences between  predictions of the ACD and SM models in low values of the compactification parameter, $1/R$.
\item We observe overall  considerable differences between  predictions of the full QCD and HQET for double lepton polarization asymmetries.
\item All polarization asymmetries have the same sign for all leptons except the $P_{TT}$, which predicts a different sign for  $\tau$ compared to the $e$
 and $\mu$. In the case of $e$
 and $\mu$ and HQET, the $P_{NN}$ changes its sign around $1/R=600~GeV$. In $P_{NN}$,  the full QCD predicts different sign for the ACD and SM models for these two leptons although the SM results are very small.
\end{itemize}

\begin{figure}
\centering
\begin{tabular}{ccc}
\epsfig{file=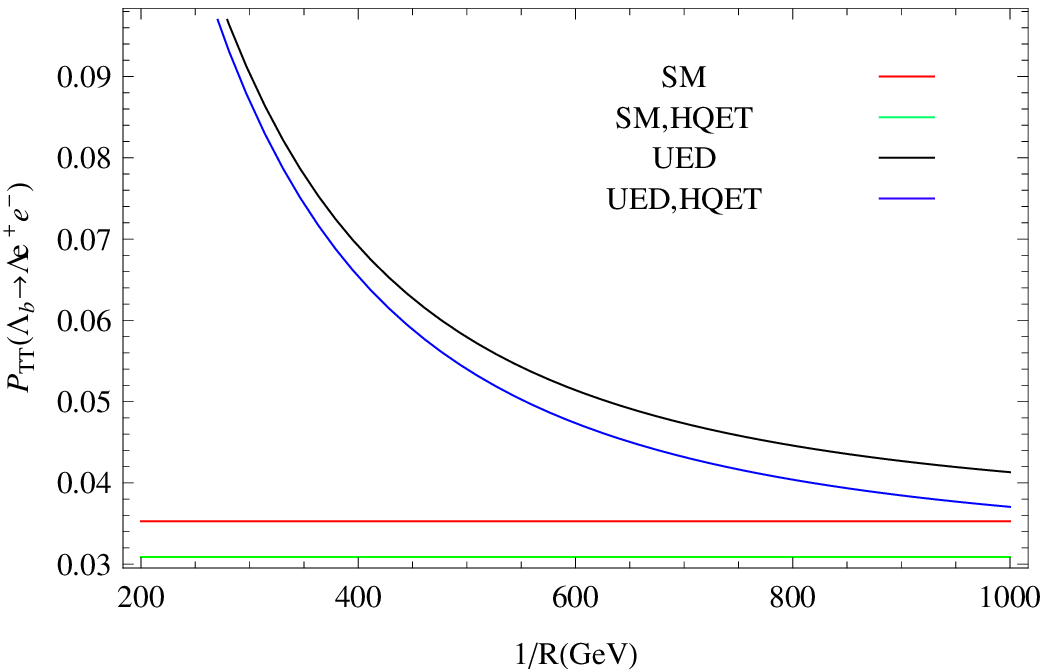,width=0.33\linewidth,clip=} &
\epsfig{file=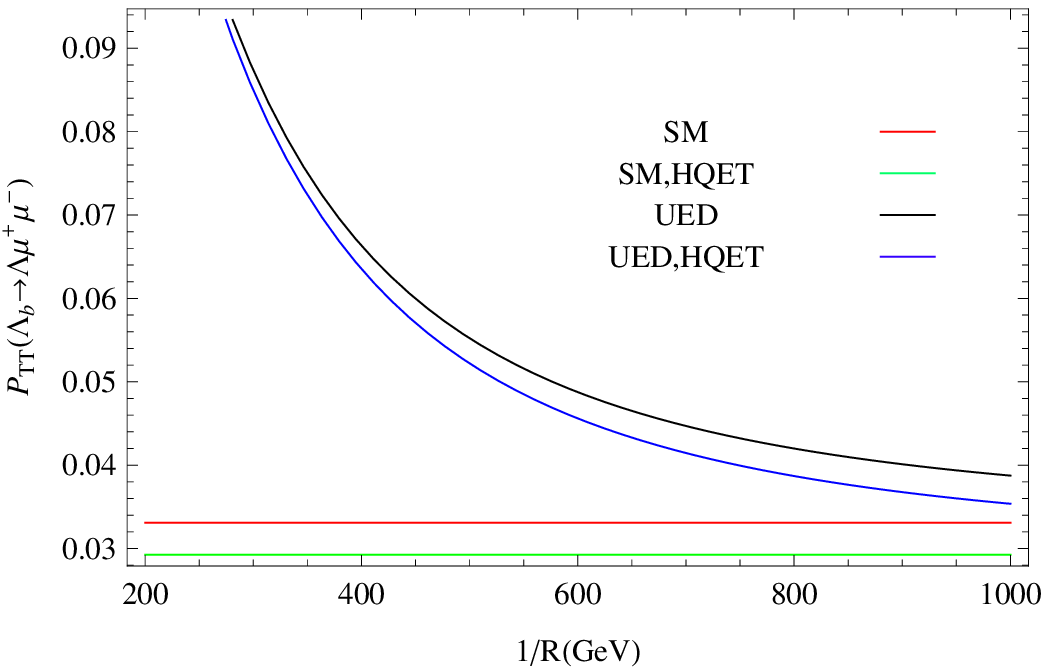,width=0.33\linewidth,clip=} &
\epsfig{file=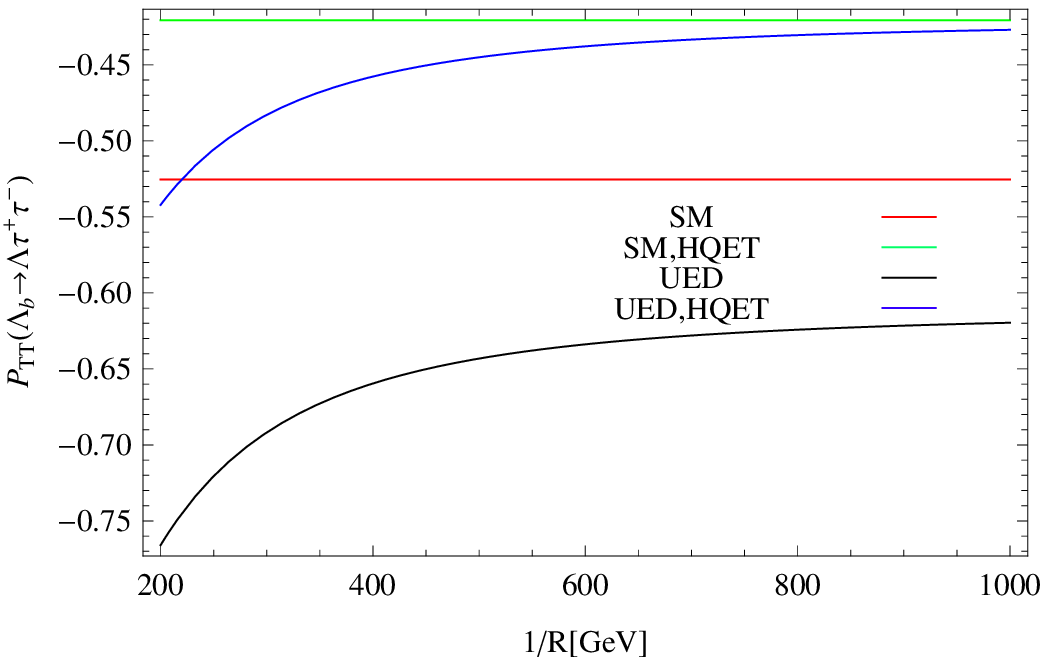,width=0.33\linewidth,clip=}
\end{tabular}
\caption{The dependence of $P_{TT}(\hat s,1/R)$  on compactification factor, $1/R$ for different leptons at $\hat s=0.5$.}
\end{figure}
%%%%

\begin{figure}
\label{fig1} \centering
\begin{tabular}{ccc}
\epsfig{file=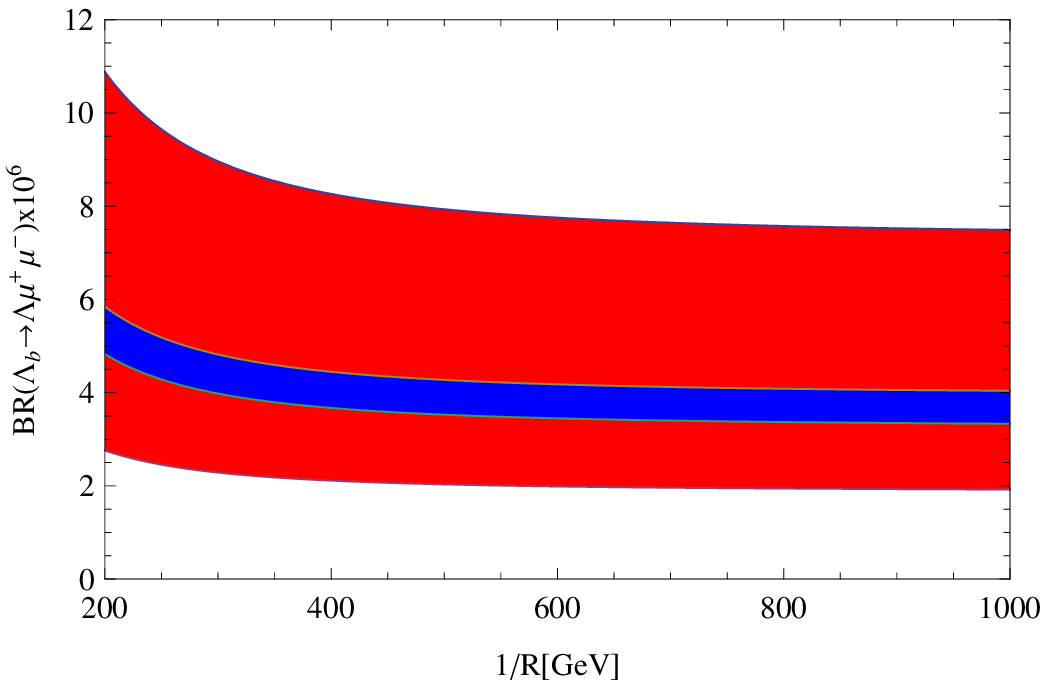,width=0.43\linewidth,clip=} &
\epsfig{file=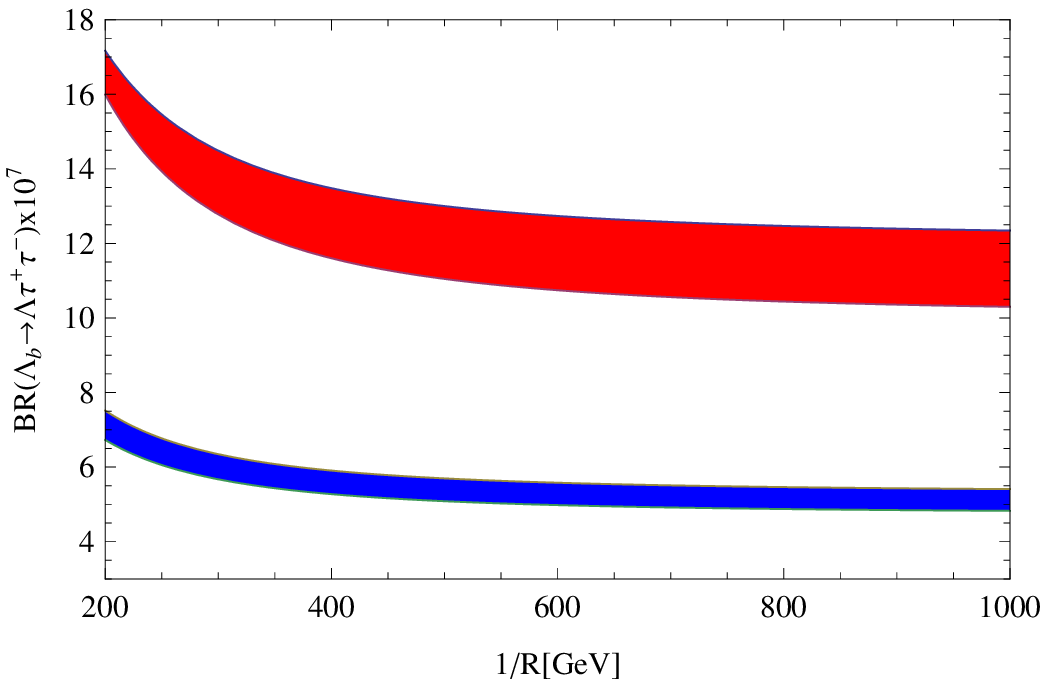,width=0.43\linewidth,clip=}
\end{tabular}
\label{nihan}
\caption{The dependence of  branching ratios on compactification
factor, $1/R$, when  errors of the form factors   are considered. The red and blue bands  belong to the full QCD and  HQET, respectively.}
\end{figure}
\begin{figure}
\centering
\begin{tabular}{ccc}
\epsfig{file=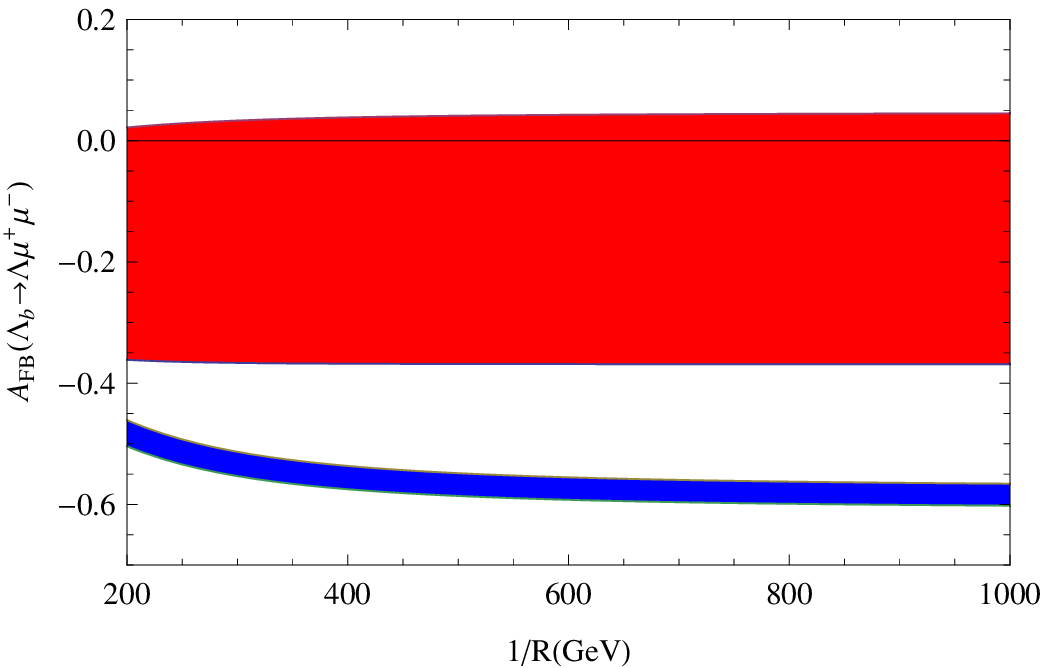,width=0.43\linewidth,clip=} &
\epsfig{file=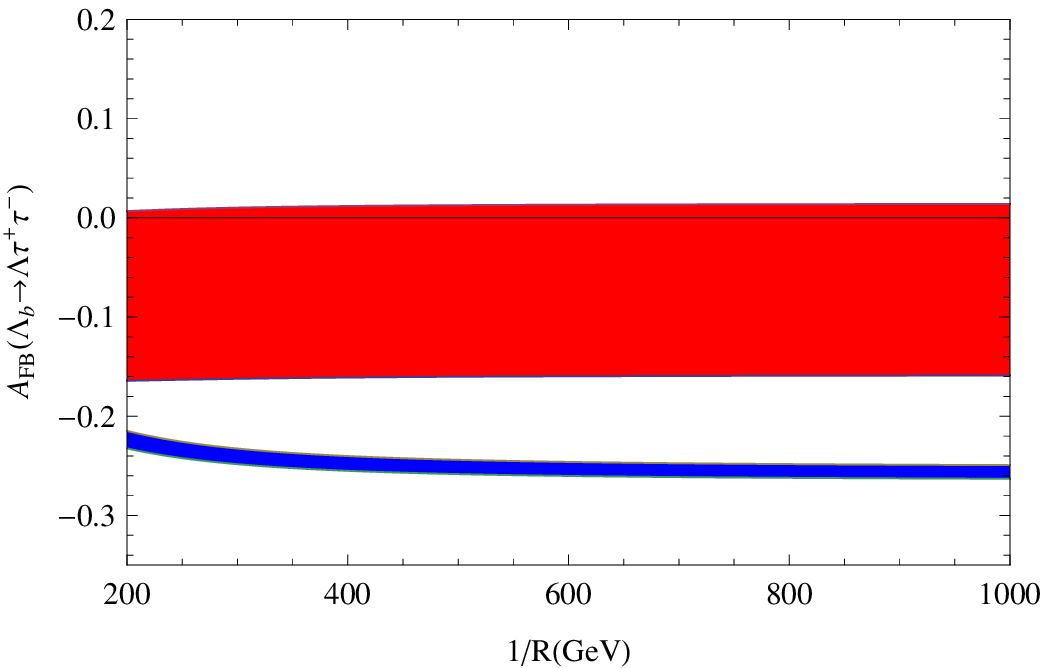,width=0.43\linewidth,clip=}
\end{tabular}
\caption{The same as Figure 10, but for  ${\cal A}_{FB}$ asymmetries.}
\end{figure}

\begin{figure}
\centering
\begin{tabular}{ccc}
\epsfig{file=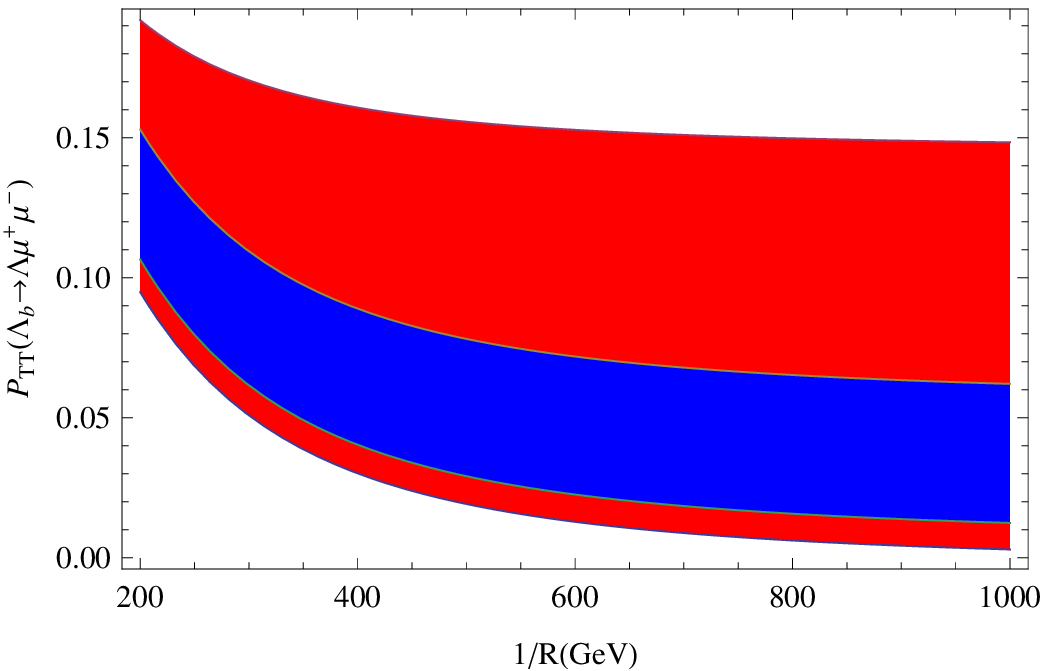,width=0.43\linewidth,clip=} &
\epsfig{file=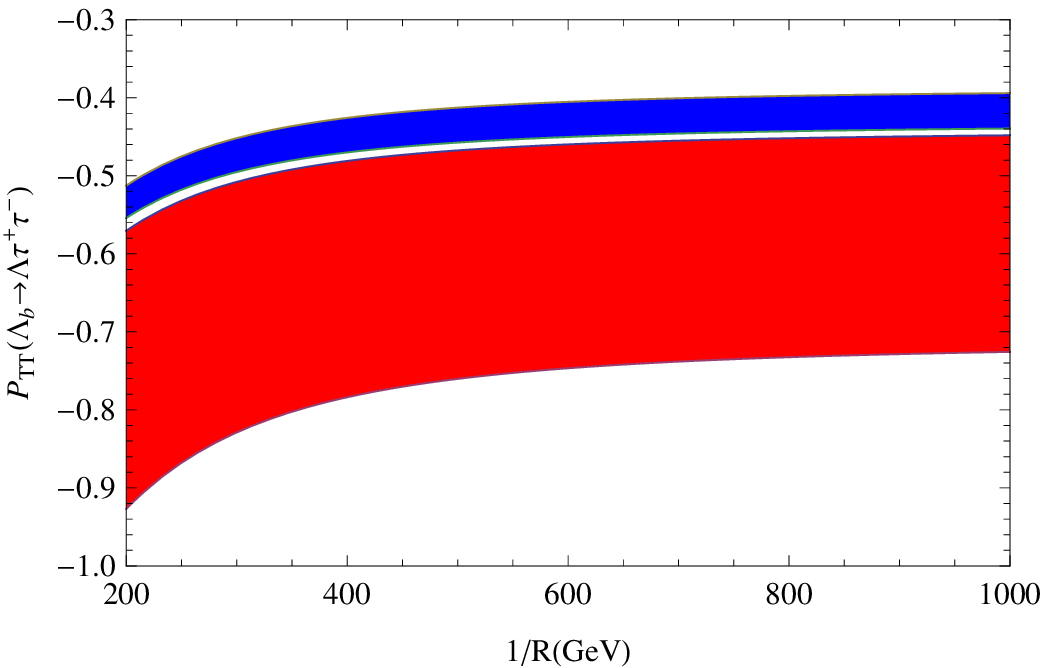,width=0.43\linewidth,clip=}
\end{tabular}
\caption{The same as Figure 10, but for $P_{TT}$ polarization.}
\end{figure}
\begin{figure}
\centering
\begin{tabular}{ccc}
\epsfig{file=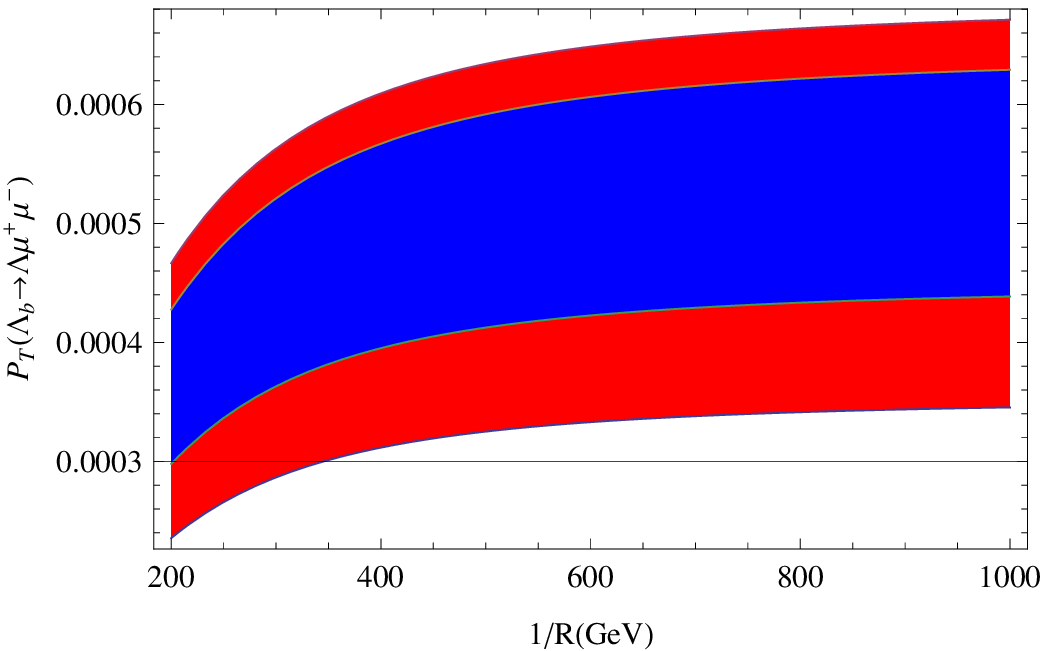,width=0.43\linewidth,clip=} &
\epsfig{file=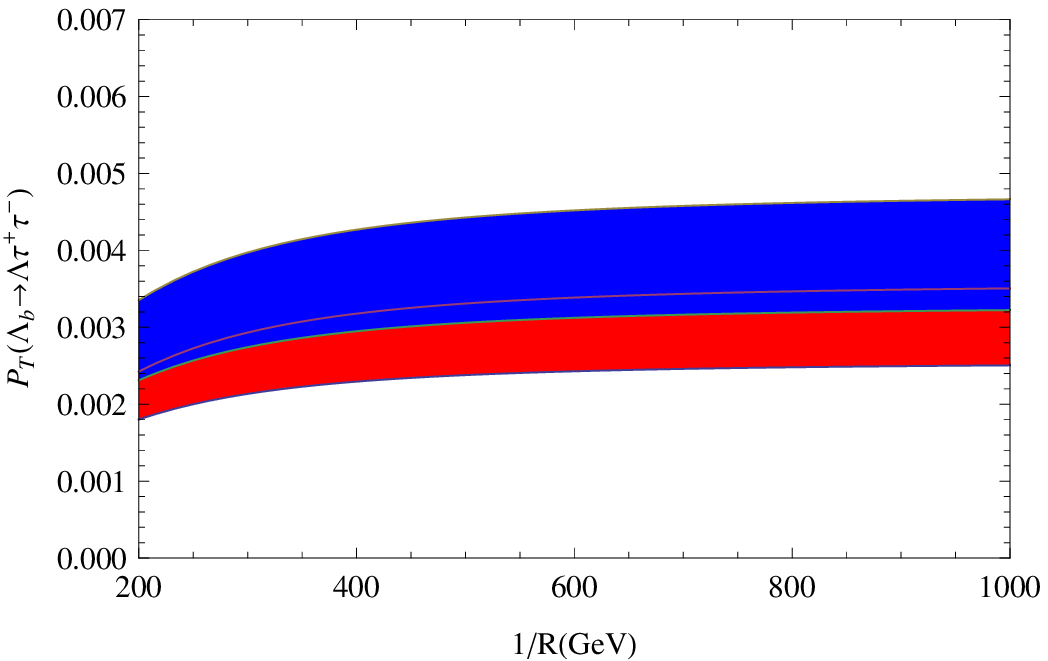,width=0.43\linewidth,clip=}
\end{tabular}
\caption{The same as Figure 10, but for $P_{T}$ polarization.}
\end{figure}
%%%%
At the end of this section, we would like to compare the full theory and HQET predictions on some observables considering the errors of form factors. In Figs. 1-9, we compared the results of two theories when the central values of 
the form 
 factors are used. Now, in Figs. 10-13, we depict the dependence of some considered observables on compactification factor, $1/R$ at a fixed value of $\hat s=0.5$ and  compare predictions of two theories  when the uncertainties of the form factors are taken into account. The red bands in these figures  belong 
to the full theory and they are obtained considering the errors of the form factors presented in \cite{kazizi}, while the blue bands correspond to the HQET and they are obtained 
using the errors of the form factors
presented in \cite{huang}. Here, we should stress that the reported errors of the form factors in HQET are small comparing those presented in full QCD, hence the HQET bands are narrow comparing to the full theory
bands. From  figure 10, we see a significant difference
 between the predictions of two theories for $\tau$ case, while in the $\mu$ case, the HQET band lies inside the full QCD region. In the case of ${\cal A}_{FB}$ in figure 11, we see also considerable difference between
delimited regions of full and HQET theories for both leptons. In the case of $\mu$ and $P_{TT}$ and $P_{T}$ polarizations (see figures 12 and 13), predictions of the HQET lie inside the full theory bands, but in the case of $\tau$ 
and  $P_{TT}$, the band of HQET
 is out of the band of full theory but very close to it. In $P_{T}$ polarization and $\tau$, two bands partly coincide with each other.

\section{Conclusion}
We analyzed the  branching
ratio, forward-backward asymmetry, double lepton polarization asymmetries and
polarization of the $\Lambda$ baryon for the channel, $\Lambda_b \rightarrow
\Lambda \ell^+ \ell^-$ in the  universal extra dimension scenario using the form factors obtained from both full QCD and HQET. For each case, we compared the obtained results with  predictions of the SM. In lower values of the compactification 
factor, we see considerable discrepancy between the UED and SM models. However, when $1/R$ grows, the results of UED tend to diminish and at $1/R=1000~GeV$, two models have approximately the same predictions.
 The order
of magnitude for branching ratios shows a possibility to study this channel at LHCb. The obtained results for the branching fractions  show also that this transition is more probable in full QCD compared to the HQET.
For other observables, we see also overall substantial differences between predictions of the full theory and HQET specially when the central values of the form factors from both theories are used. 
Any measurements on the considered physical
 quantities in this manuscript  and their comparison with our predictions, can give useful information about existing of extra dimensions.


\begin{thebibliography}{99}
 \bibitem{Antoniadis1} I. Antoniadis, Phys. Lett. B 246, 377 (1990).
\bibitem{Antoniadis2} I. Antoniadis, N.
Arkani-Hamed, S. Dimopoulos and G. Dvali, Phys. Lett. B 439, 257
(1998).
\bibitem{arkani} N. Arkani-Hamed, S. Dimopoulos and G. Dvali, Phys. Lett. B 429, 263
(1998); Phys. Rev. D 59, 086004 (1999). 


\bibitem{acdd} T. Appelquist, H. C. Cheng and B. A. Dobrescu,
  Phys. Rev.  D 64, 035002 (2001).
\bibitem{Buras:2002ej}
  A.~J.~Buras, M.~Spranger and A.~Weiler,
  %``The impact of universal extra dimensions on the unitarity triangle and rare
  %K and B decays. ((U)),''
  Nucl.\ Phys.\ B  660, 225 (2003).
%  [arXiv:hep-ph/0212143].
  %%CITATION = HEP-PH 0212143;%%
\bibitem{R7623}
  A. J. Buras, A. Poschenrieder, M. Spranger
  Nucl. Phys. B  D 678, 455 (2004).
\bibitem{R7624}
  P. Colangelo, F. De Fazio, R. Ferrandes, T. N. Pham, Phys. Rev. D7 3 (2006) 115006.
\bibitem{10susy} G. Buchalla, G. Hiller and G. Isidori,
  Phys. Rev. D 63 (2000) 014015.

\bibitem{11darkmatter} C. Bird, P. Jackson, R. Kowalewski, M. Pospelov,
  Phys. Rev. Lett. 93 (2004) 201803.
\bibitem{T. Appelquist} T. Appelquist and H. U. Yee, Phys. Rev. D 67 (2003) 055002.

%\bibitem{ikiuc} K. Agashe, N. G. Deshpande,  G. H. Wu, Phys. Lett. B 514 (2001) 309; B 511 (2001) 85; T. Appelquist,  B. A. Dobrescu, Phys. Lett. B 516 (2001) 85.



\bibitem{kazizi} T. M. Aliev, K. Azizi, M. Savci, Phys. Rev. D, 81, 056006, (2010).
 \bibitem{huang} C. S. Huang, H. G. Yan, Phys. Rev. D 59, 114022 (1999).

\bibitem{colangelo}  P. Colangelo, F. De Fazio, R. Ferrandes, T. N. Pham, Phys. Rev. D 77, 055019 (2008).
 \bibitem{R7601} T. M. Aliev, M. Savc{\i}, Eur. Phys. J. C 50, 91 (2007).
  \bibitem{wang} Yu-Ming Wang, M. Jamil Aslam, Cai-Dian Lu, Eur. Phys. J. C 59, 847 (2009).  





 \bibitem{colangelobey} M.V. Carlucci, P. Colangelo, F. De Fazio, Phys. Rev. D 80, 055023 (2009). 

  \bibitem{Ishtiaq}  Ishtiaq Ahmed, M. Ali Paracha, M. Jamil Aslam, Eur. Phys. J. C 54, 591 (2008).
   \bibitem{Asif} Asif Saddique, M. Jamil Aslam, Cai-Dian Lu, Eur. Phys. J. C 56, 267 (2008). 
           \bibitem{Zeynali}       V. Bashiry, K. Zeynali, Phys. Rev. D 79, 033006 (2009). 
%\bibitem{Georgi}H.~Georgi, A.~K.~Grant and G.~Hailu, Phys.\ Rev.\ D {\bf 63}
%(2001) 064027, [arXiv:hep-ph/0007350].
\bibitem{R7626}
  A. Buras, M. Misiak, M. M\"{u}nz and S. Pokorski,
  Nucl. Phys. B 424, 374 (1994).

\bibitem{R7627}
  M. Misiak,
  Nucl. Phys. B  393, 23 (1993);
  Erratum ibid  B  439, 161 (1995).

\bibitem{R762777}
    B. Buras, M. M\"{u}nz,
  Phys. Rev. D  52, 186 (1995).

\bibitem{27alievozpineci}  T. M. Aliev, A. Ozpineci, M. Savci,
  Phys. Rev. D 65 (2002) 115002.
\bibitem{28Mannel} T. Mannel, W. Roberts and Z. Ryzak,
  Nucl. Phys. B355 (1991) 38.
\bibitem{chen} C. H. Chen,  C. Q. Geng, Phys. Rev. D 63 (2001) 054005;
  Phys. Rev. D 63 (2001) 114024; Phys. Rev. D 64 (2001) 074001.
  %\bibitem{R7632}
  %CDF Collaboration,
  %prep: hep--ex/0507091.



\bibitem{30ozpineci} T. M. Aliev, A. Ozpineci, M. Savci, C. Yuce ,
  Phys. Lett. B 542 (2002) 229.
\bibitem{savcibey} T. M. Aliev, A. Ozpineci, M. Savci, Nucl .Phys. B 649 (2003) 168.
\bibitem{Giri}  A. K. Giri, R. Mohanta, Eur. Phys. J. C 45, 151 (2006).

\bibitem{33Yao:2006px} C. Amsler et al. [Particle Data Group],
  Phys. Lett. B 667, 1 (2008).
\bibitem{R7615}
  T. M. Aliev, A. \"{O}zpineci,  M. Savc{\i},
  Phys. Rev. D  67, 035007 (2003).

\bibitem{ozpinecibey} T. M. Aliev, A. Ozpineci, M. Savci, arXiv:hep-ph/0301019.
%\bibitem{R7630}
 % T. Mannel, W. Roberts, and Z. Ryzak,
  %Nucl. Phys. B  355, 38 (1991).
 %\bibitem{R7631}
  %S. Eidelman {\it et al.}, Particle data Group,
  %Phys. Lett. B  592, 1 (2004).

\bibitem{vahli} T. M. Aliev, V. Bashiry, M. Savci, Eur. Phys. J. C 38 (2004) 283.
\bibitem{0608143} T. M. Aliev, M. Savci, B. B. Sirvanli, Eur. Phys.J. C 52, 375 (2007).
  \bibitem{R7710}
  W. Bensalem, D. London, N. Sinha and R. Sinha,
  Phys. Rev. D  67, 034007 (2003).


\end{thebibliography}
\end{document}